\documentclass[aps,prd,amssymb,amsmath,amsfonts,superscriptaddress,nofootinbib,eqsecnum,reprint,showpacs,longbibliography]{revtex4-1}

\usepackage{graphicx}
\usepackage{lmodern}
\usepackage{amsmath,amssymb}
\usepackage{mathrsfs}
\usepackage{amsfonts}
\usepackage[utf8]{inputenc}
\usepackage{url}
\usepackage[colorlinks]{hyperref}
\usepackage{xcolor}
\usepackage{multirow}
\usepackage[normalem]{ulem}
\usepackage{orcidlink}
\usepackage{placeins}
\usepackage{mathtools}

\definecolor{dodgerblue}{HTML}{1E90FF}
\definecolor{viennared}{HTML}{DA0A14}
\definecolor{ctorange}{HTML}{FF6C0C}
\definecolor{wales}{HTML}{ff0038}
\definecolor{benettongreen}{HTML}{009421}
\definecolor{ferrarired}{HTML}{ff2800}
\definecolor{austriawienpurple}{HTML}{441678}
\definecolor{steiermarkgruen}{HTML}{006747}

\hypersetup{
     colorlinks=true,
     linkcolor=dodgerblue,
     filecolor=dodgerblue,
     citecolor = dodgerblue,      
     urlcolor=dodgerblue,
     }

\newcommand{\Birmingham}{School of Physics and Astronomy and Institute for Gravitational Wave Astronomy, University of Birmingham, Edgbaston, Birmingham, B15 2TT, United Kingdom}

\newcommand{\Turin}{INFN Sezione di Torino, Via P. Giuria 1, 10125 Torino, Italy}
\newcommand{\IHES}{Institut des Hautes Etudes Scientifiques, 91440 Bures-sur-Yvette, France}

\DeclarePairedDelimiterX{\norm}[1]{\lVert}{\rVert}{#1}

\begin{document}

\title{Strong-field scattering of two spinning black holes:  Numerical Relativity versus post-Minkowskian gravity}

\author{Piero Rettegno \orcidlink{0000-0001-8088-3517}} 
\email{piero.rettegno@to.infn.it}
\affiliation{\Turin} 
\affiliation{\Birmingham}

\author{Geraint Pratten \orcidlink{0000-0003-4984-0775}} 
\email{g.pratten@bham.ac.uk}
\affiliation{\Birmingham}

\author{Lucy M. Thomas \orcidlink{0000-0003-3271-6436}} 
\email{lthomas@star.sr.bham.ac.uk}
\affiliation{\Birmingham}

\author{Patricia Schmidt \orcidlink{0000-0003-1542-1791}} 
\email{p.schmidt@bham.ac.uk}
\affiliation{\Birmingham}

\author{Thibault Damour} 
\email{damour@ihes.fr}\affiliation{\IHES}

\date{\today}

\begin{abstract}

 Highly accurate models of the gravitational-wave signal from coalescing compact binaries are built by completing analytical computations of the binary dynamics with non-perturbative information from numerical relativity (NR) simulations.  In this paper we present four sets of NR simulations of equal-mass black hole binaries that undergo strong-field scattering:
(i) we reproduce and extend the nonspinning simulations first presented in [Damour \textit{et al.}, Phys.Rev.D 89 (2014) 8, 081503], (ii) we compute two suites of nonspinning simulations at higher energies, probing stronger field interactions, (iii) we present a series of \textit{spinning} simulations including, for the first time, unequal-spin configurations.
When comparing the NR scattering angles to analytical predictions based on state-of-the-art post-Minkowskian (PM) calculations, we find that PM-expanded scattering angles show poor convergence towards NR data.
By contrast, a resummed computation of scattering angles via a spin-dependent, radiation-reacted, effective-one-body potential shows excellent agreement for both nonspinning and spinning configurations.

\end{abstract}

\maketitle 

\section{Introduction}
\label{sec:intro}

The detection and characterisation of gravitational-wave (GW) observations~\cite{LIGOScientific:2018mvr,LIGOScientific:2020ibl,LIGOScientific:2021djp, Nitz:2021zwj, Olsen:2022pin} from compact binary coalescences relies on theoretical predictions of the emitted signal. Highly accurate models of coalescing compact binaries are a crucial requirement to perform precise measurements of the properties of black holes (BHs) and neutron stars, to determine their underlying astrophysical distributions and to perform tests of General Relativity in the strong-field regime.
Such waveform models are generally built using both analytical and numerical approximations to Einstein's equations for a binary system in quasi-circular orbits.

Numerical Relativity (NR) simulations of quasi-circular binaries~\cite{Pretorius:2005gq, Campanelli:2005dd,Mroue:2013xna,Husa:2015iqa,Jani:2016wkt, Boyle:2019kee,Healy:2019jyf,Hamilton:2023qkv} provide the most accurate representation of the emitted waveform, especially during the plunge, merger and ringdown phases. Such simulations can be used to build NR surrogates~\cite{Field:2013cfa,Blackman:2017dfb, Varma:2019csw}, whose main limitation is their restricted parameter space coverage.

Semi-analytical GW approximants, such as effective-one-body (EOB)~\cite{Buonanno:1998gg,Buonanno:2000ef,Gamba:2021ydi,Ossokine:2020kjp,Ramos-Buades:2023ehm} and phenomenological models~\cite{Hannam:2013oca,Pratten:2020fqn,Garcia-Quiros:2020qpx,Pratten:2020ceb,Hamilton:2021pkf}, use a combination of numerical and analytical information. These approximants generally make use of the post-Newtonian (PN) expansion~\cite{Blanchet:1989ki,Blanchet:2013haa,Damour:2014jta,Levi:2015uxa,Bini:2017wfr,Schafer:2018kuf,
Bini:2019nra,Bini:2020wpo,Bini:2020nsb,Bini:2020hmy,Antonelli:2020ybz,Blumlein:2020pyo,Blumlein:2021txe,Mandal:2022nty,Mandal:2022ufb}, which assumes small  
velocities ($v/c \ll 1$) together with weak fields  [$G M / (r c^2) \ll 1$], making it apt to describe quasi-circular inspirals. However, with the improving sensitivity of current detectors and ongoing development of a third generation of GW detectors~\cite{Reitze:2019iox,Punturo:2010zz,LISA:2017pwj},
the possibility of detecting eccentric binaries and hyperbolic encounters has risen in prominence~\cite{Romero-Shaw:2020thy,CalderonBustillo:2020odh,Gayathri:2020coq,Gamba:2021gap}. The detection and characterization of binaries on non-circular orbits is thought to be a powerful tracer for constraining the formation and evolution channels of astrophysical BHs~\cite{OLeary:2005vqo,OLeary:2008myb,Samsing:2013kua,Rodriguez:2016kxx,Belczynski:2016obo,Samsing:2017xmd,Romero-Shaw:2019itr,Zevin:2021rtf}.

Analytical models of non-circular binaries cannot solely rely on the PN approximation, as the two bodies could already possess high velocities at large separations. 
In these situations, the post-Minkowskian (PM) approximation~\cite{Damour:2016gwp,Damour:2017zjx}, which only assumes weak fields [$G M / (r c^2) \ll 1$] while allowing for large velocities, is more appropriate.
PM contributions have been computed using various approaches to gravitational scattering,
such as: scattering amplitudes (see, e.g.,~\cite{Cheung:2018wkq,Guevara:2018wpp,Kosower:2018adc,Bern:2019nnu,
Bern:2019crd,Bern:2019nnu,Bjerrum-Bohr:2019kec,Herrmann:2021tct,
Bern:2021dqo,Bern:2021yeh,Bjerrum-Bohr:2021din,Manohar:2022dea,
Saketh:2021sri}); eikonalization (e.g.,~\cite{KoemansCollado:2019ggb,DiVecchia:2019kta,DiVecchia:2021bdo,DiVecchia:2022nna}); effective field theory (e.g.,~\cite{Kalin:2020mvi,Kalin:2020fhe,
Mougiakakos:2021ckm,Dlapa:2021npj,Dlapa:2021vgp,Kalin:2022hph,Dlapa:2022lmu}); and worldline (classical
or quantum) field theory (e.g.,~\cite{Mogull:2020sak,Riva:2021vnj,Jakobsen:2021smu,Jakobsen:2022psy,Bini:2021gat,Bini:2022wrq,Bini:2022enm}).
The resulting terms can be incorporated into semi-analytic models to improve their accuracy for binaries on eccentric and hyperbolic orbits~\cite{Chiaramello:2020ehz,Nagar:2021gss,Placidi:2021rkh,Khalil:2021txt,Ramos-Buades:2021adz,Nagar:2021xnh,Khalil:2022ylj}. 
At the same time, numerical results are necessary to determine the accuracy and validity of these approximations.
Whilst a large suite of numerical simulations of binary black holes (BBHs) on eccentric orbits have been performed~\cite{Hinder:2008kv,Gold:2011df,Lewis:2016lgx,Ramos-Buades:2019uvh,Huerta:2019oxn,Habib:2019cui,Gayathri:2020coq,Islam:2021mha,Ramos-Buades:2022lgf}, there are comparatively few simulations regarding BBH scattering ~\cite{Shibata:2008rq,Sperhake:2009jz,Damour:2014afa,Hopper:2022rwo,Healy:2023wyr}.

The aims of this paper are:
(i) to extend  the results of Ref.~\cite{Damour:2014afa} by performing higher-initial-energy NR simulations of equal-mass nonspinning BBH scattering\footnote{As a check, we have also performed simulations with the same initial energy as \cite{Damour:2014afa}.};
(ii) to perform NR simulations of the scattering of equal-mass {\it spinning} BBHs, both for equal and unequal spins, aligned with the angular momentum;
and 
(iii) to compare the so-obtained strong-field numerical scattering angles to analytical predictions based on state-of-the-art PM results\footnote{We do not compare here numerical scattering angles to state-of-the-art {\it post-Newtonian} predictions. Reference~\cite{Damour:2014afa} made such comparisons in the nonspinning case and found that PN-expanded angles performed badly, compared to corresponding PN-based EOB-defined ones. 
}.

We denote the masses of the two objects as $m_1$ and $m_2$, and the individual dimensionless spins as $\chi_i = S_i/(G m_i^2) = a_i/(G m_i)$, $i=1,2$.  In addition, we denote $M = m_1 + m_2$, $\mu = {m_1 m_2}/{(m_1+m_2)}$,
$\nu= {\mu}/{M}= {m_1 m_2}/{(m_1+m_2)^2}$. 
All our simulations will have $m_1=m_2$, so that $\nu=1/4$. 
Throughout this paper, we denote the {\it canonical} orbital angular momentum simply as $L$. It is related to the total Arnowitt-Deser-Misner (ADM) angular momentum of the system, $J$, by 
\begin{equation}
    J = L + S_1 + S_2 \, .
\end{equation}
Unless otherwise stated, we take $G = c = 1$ and generally work with dimensionless quantities.

\section{Numerical Relativity (NR) Simulations}
\label{sec:nrsim}
We performed several sequences of nonspinning and spin-aligned NR simulations modelling the scattering of two equal mass BHs. The numerical simulations were performed using the \texttt{Einstein Toolkit} (ETK)~\cite{EinsteinToolkit:2022_11}, an open source numerical relativity code built on the Cactus framework. The numerical setup of our simulations is broadly similar to that used in \cite{Damour:2014afa}, though we present the details here for completeness. 

We use Bowen-York initial data \cite{Bowen:1980yu,Brandt:1997tf} computed using the \texttt{TwoPunctures} thorn \cite{Ansorg:2004ds}. As in \cite{Damour:2014afa}, the BHs are initially placed on the $x$-axis at a separation of $\pm X$ and with initial ADM linear momenta
\begin{align}
\vec{P} = (P_x,P_y,P_z) = \pm \, P_{\rm ADM} \left(\hspace{-0.1cm}-\sqrt{1\hspace{-0.05cm}-\hspace{-0.05cm}\left(\frac{b_{\rm NR}}{2 X} \right)^2 }, \frac{{b}_{\rm NR}}{2 X}, 0 \right).
\end{align}
\newline 
where $P_{\rm ADM} = |\vec{P}|$ and $b_{\rm NR}$ denotes an impact parameter. As in \cite{Damour:2014afa}, we use $X = 50 M$. The
Bowen-York initial data determine both the initial
total ADM energy  $E^{\rm ADM}_{\rm in} $ of the system, and the  total initial ADM angular momentum of the system. The latter is given by the simple formula
\begin{align}
    J^{\rm ADM}_{\rm in} =  L_{\rm in} + S_1+S_2\,,
\end{align}
where the initial canonical orbital angular momentum $L_{\rm in}$ is related to 
$b_{\rm NR}$ and $P_{\rm ADM}$ by~\cite{Damour:2014afa}
\begin{align}
    L_{\rm in} = 2 X |P_y| = \, P_{\rm ADM} \, b_{\rm NR}\,.
\end{align}
We adopt as initial lapse profile $\alpha = \psi^{-2}_{\rm BL}$, where $\psi_{\rm BL}$ denotes the Brill-Lindquist conformal factor~\cite{ Brandt:1997tf,Alcubierre:2002kk,Campanelli:2005dd}. In Appendix~\ref{app:nrerrors}, we present a check that the uncertainty linked to the choice of initial lapse profile has a subdominant effect  on the inferred scattering angle $\theta_{\rm NR}$ compared to the errors in the polynomial fits used to measure the scattering angle.

Time evolution is performed using the $W$-variant \cite{Marronetti:2007wz} of the BSSNOK formulation \citep{Shibata:1995we,Baumgarte:1998te,Nakamura:1987zz} of the Einstein field equations as implemented by the \texttt{McLachlan} \cite{Brown:2008sb} thorn. We evolve the BHs using moving punctures gauge conditions \cite{Baker:2005vv,Campanelli:2005dd}, the lapse is evolved using the $1+\log$ condition \cite{Bona:1994dr}, and the shift is evolved using the hyperbolic $\tilde{\Gamma}$-driver equation \cite{Alcubierre:2002kk}. We use eight-order accurate finite differencing stencils with Kreiss-Oliger dissipation \cite{Kreiss:1973}. Adaptive mesh refinement is provided by \texttt{Carpet}, with the near zone being computed with high-resolution Cartesian grids that track the motion of the BHs and the wave extraction zone being computed on spherical grids using the \texttt{Llama} multipatch infrastructure \cite{Pollney:2009yz}. The apparent horizons are computed using \texttt{AHFinderDirect} \cite{Thornburg:2003sf} and the spin angular momenta are calculated using the dynamical horizon formalism provided by the \texttt{QuasiLocalMeasures} thorn \cite{Dreyer:2002mx}.  

Similarly to \cite{Damour:2014afa}, we found that the total energy and angular momentum of the system left after the release of the burst of spurious radiation present in the initial data differ from the
corresponding ADM quantities computed from the initial data only by negligible fractions of order $10^{-5}$.

All ETK simulations were managed using \texttt{Simulation Factory} \cite{simfactory} and post-processing of NR data has made used of the open source Mathematica package \texttt{Simulation Tools} \cite{SimulationTools}. 

\subsection{Extracting the Scattering Angle}
In order to calculate the scattering angle, we follow the prescription detailed in \cite{Damour:2014afa}. The motion of the BHs is tracked by the Cartesian coordinates of the punctures in the center-of-mass frame. We convert the tracks to polar coordinates $(r,\varphi)$ (in the $x-y$ plane of motion) for each BH and treat the incoming $\varphi_{{\rm in},i} (r)$ and outgoing $\varphi_{{\rm out},i} (r)$ paths for the $i$-th BH separately. We fit each of the paths to a polynomial of order $n$ in terms of $u = 1/r$ and extrapolate to find the asymptotic angle $\varphi^{\infty}_{{\rm in/out},i}$. The resulting scattering angle for the $i$-th particle is given by 
\begin{align}
    \theta_i = \varphi^{\infty}_{{\rm out},i} - \varphi^{\infty}_{{\rm in},i} - \pi \,. 
\end{align}
\newline 
Our choice of fitting window for $(u,\varphi)$ follows  \cite{Damour:2014afa}, and we assume $r \in[14,80] M$ and $r \in [20,100]M$ for the incoming and outgoing trajectories respectively. We implement a least-squares fitting method that uses a singular value decomposition (SVD) to drop singular values smaller than $10^{-13}$ times the maximum singular value, as was used in \cite{Damour:2014afa}. In order to gauge the errors on the scattering angle, we perform a number of sanity checks. The first check is to gauge the impact of the polynomial order $n$ on the resulting scattering angle. As in \cite{Damour:2014afa}, the preferred polynomial order is taken to be the lowest order for which the SVD method allows for variation in the constant term. The extrapolation error is then estimated by finding the maximum and minimum scattering angle inferred over all polynomial orders between $2$ and $n-1$. This generally leads to  dissymmetric error bounds on the scattering angle:  $\theta^{+ \delta_+}_{- \delta_-}$.
A value of $0$ in $\delta_+$ or $\delta_-$ occurs when the preferred polynomial is setting the bound. More precisely, $\delta_- =0 $ (respectively $\delta_+ =0 $) occurs when the lower order polynomial fits over-estimate (respectively, under-estimate) the scattering angle relative to the preferred order. When least-square fitting our numerical results to analytical templates, we shall use a symmetrized version of the error bounds, namely $\pm \frac12(\delta_+ + \delta_-)$.

Other sanity checks on the robustness of the derived scattering angle include testing the impact of the fitting window, testing the resolution of our numerical simulations, and varying gauge choices in the numerical evolution. We find that the inferred errors on the scattering angle remain subdominant to the choice of polynomial order, in agreement with the conclusions in \cite{Damour:2014afa}. See Appendix~\ref{app:nrerrors} for further details.  

\subsection{Nonspinning Scattering Simulations}

Though the main aim of the present work is to simulate the scattering of spinning BBHs, we also performed systematic sequences of nonspinning simulations of BBHs with fixed initial energy and varying initial angular momenta in order to be able to extract direct numerical estimates of the EOB interaction potential (see Sec.~\ref{sec:NRPM} below).
For the first sequence, we reproduced and extended the results of Ref.~\cite{Damour:2014afa}, with an initial (adimensionalized) ADM energy $\hat{E}^{\rm ADM}_{\rm in} \equiv E_{\rm in}^{\rm ADM} / M$ equal to
$\hat{E}^{\rm ADM}_{\rm in, 1} \simeq 1.02264$ and varying initial (rescaled) orbital angular momentum $\hat{L}_{\rm in} \equiv L_{\rm in} / M^2$.
This allowed us both to cross-check the accuracy of our computations and to probe more precisely the boundary between scattering and plunge. 
We also performed two more sequences at the higher incoming energies, $\hat{E}_{\rm in, 2}^{\rm ADM} \simeq 1.04033$ and $\hat{E}_{\rm in, 3}^{\rm ADM} \simeq 1.05548$. 
These explore a stronger-field region of the parameter space, with NR impact parameters going down to $\hat{b}_{\rm NR} \equiv b_{\rm NR}/M \simeq 6$ (with
corresponding minimum isotropic EOB radial coordinate $\bar{r}^{\rm min} \simeq 2$, see below). 

\begin{figure}[t]
	\includegraphics[width=0.48\textwidth]{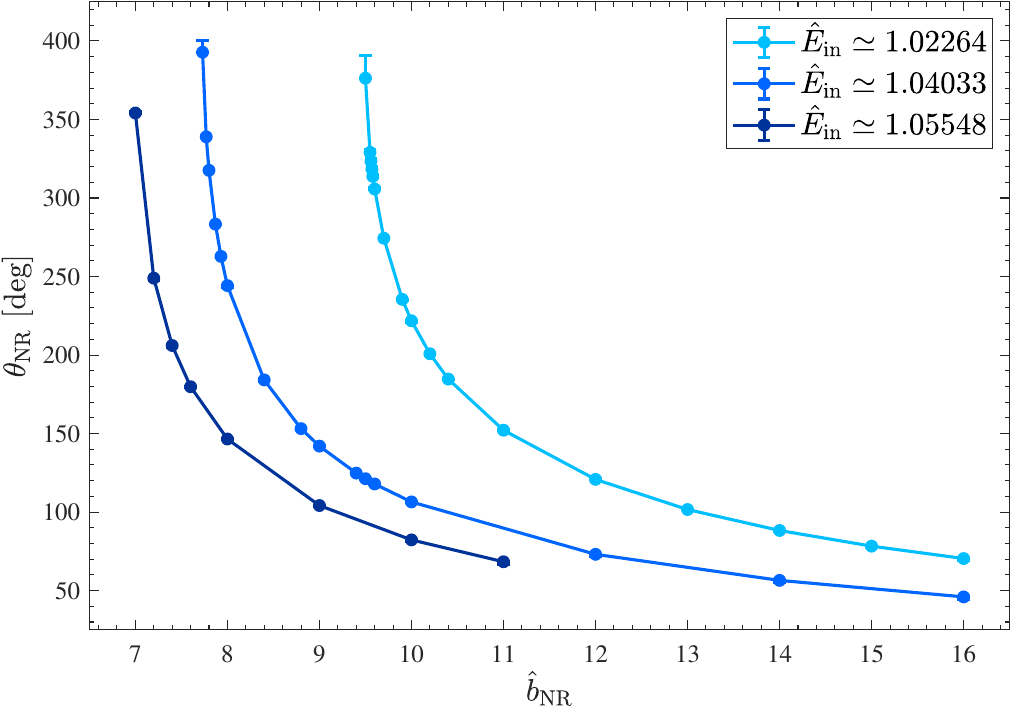}
	\caption{
    \label{fig:chiNR_nospin}
	NR scattering angles against impact parameter for the (equal-mass) nonspinning simulations.
    Higher-energy systems are able to probe stronger fields (smaller separations) without plunging.
	}
\end{figure}

We report the results of all our nonspinning simulations in Table~\ref{tab:NRnospin}. 
In Fig.~\ref{fig:chiNR_nospin} we show the scattering angles versus the impact parameter $\hat{b}_{\rm NR}$ for all three series of nonspinning systems. For the nonspinning simulations, we have that $\theta_{\rm NR} = \theta_{1}= \theta_{2}$. 

\begin{table}[htbp]
    \caption{
    \label{tab:NRnospin}
    Equal-mass, nonspinning NR simulations.   
    The scattering angles have been replaced by dots where the BHs have plunged. 
    An asterisk denotes simulations where there is uncertainty on an eventual plunge. 
    Longer simulations are required to determine whether the system becomes bound after the first encounter. 
    }
    \setlength{\tabcolsep}{0.4cm}
    \begin{tabular}{c c c c}
    \hline
    \hline   
    $\hat{b}_{\rm NR}$ & $\hat{E}_{\rm in}^{\rm{ADM}}$ & $\hat{L}_{\rm in}^{\rm{ADM}}$ & $\theta_{\rm{NR}} \left[\rm deg\right]$ \\
    \hline
        9.40 & 1.02264 & 1.07690 & $\cdots$ \\
        9.50 & 1.02264 & 1.08840 & $\phantom{\ast} 376.275^{+0.026}_{-14.69} \ast$ \\
        9.55 & 1.02264 & 1.09410 & $329.057^{+0.003}_{-1.534}$ \\
        9.56 & 1.02264 & 1.09520 & $323.422^{+0.000}_{-1.914}$ \\
        9.57 & 1.02264 & 1.09640 & $318.394^{+0.000}_{-1.575}$ \\
        9.58 & 1.02264 & 1.09750 & $313.764^{+0.000}_{-1.331}$ \\
        9.60 & 1.02264 & 1.09980 & $305.734^{+0.056}_{-0.694}$ \\
        9.70 & 1.02264 & 1.11130 & $274.368^{+0.074}_{-0.016}$ \\
        9.90 & 1.02264 & 1.13420 & $235.447^{+0.912}_{-0.003}$ \\
        10.00 & 1.02264 & 1.14560 & $221.823^{+0.762}_{-0.002}$ \\
        10.20 & 1.02264 & 1.16860 & $200.810^{+0.620}_{-0.004}$ \\
        10.40 & 1.02264 & 1.19150 & $184.684^{+0.221}_{-0.002}$ \\
        11.00 & 1.02264 & 1.26020 & $152.106^{+0.055}_{-0.446}$ \\
        12.00 & 1.02264 & 1.37480 & $120.804^{+0.013}_{-0.307}$ \\
        13.00 & 1.02264 & 1.48930 & $101.616^{+0.059}_{-0.002}$ \\
        14.00 & 1.02264 & 1.60390 & $88.260^{+0.337}_{-0.002}$ \\
        15.00 & 1.02264 & 1.71850 & $78.296^{+0.520}_{-0.002}$ \\
        16.00 & 1.02264 & 1.83300 & $70.404^{+0.927}_{-0.003}$ \\
        \hline
        7.67 & 1.04032 & 1.15050 & $\cdots$ \\
        7.73 & 1.04032 & 1.15950 & $\phantom{\ast} 392.815^{+0.006}_{-7.477} \ast$ \\
        7.77 & 1.04032 & 1.16550 & $338.973^{+0.156}_{-0.756}$ \\
        7.80 & 1.04032 & 1.17000 & $317.637^{+0.142}_{-0.444}$ \\
        7.87 & 1.04032 & 1.18050 & $283.359^{+0.343}_{-0.007}$ \\
        7.93 & 1.04032 & 1.18950 & $262.825^{+0.749}_{-0.008}$ \\
        8.00 & 1.04032 & 1.20000 & $244.210^{+1.220}_{-0.005}$ \\
        8.40 & 1.04033 & 1.26000 & $184.138^{+0.439}_{-0.004}$ \\
        8.80 & 1.04033 & 1.32000 & $153.119^{+0.226}_{-0.227}$ \\
        9.00 & 1.04033 & 1.35000 & $141.986^{+0.244}_{-0.213}$ \\
        9.40 & 1.04033 & 1.41000 & $124.805^{+0.154}_{-0.238}$ \\
        9.50 & 1.04033 & 1.42500 & $121.233^{+0.180}_{-0.153}$ \\
        9.60 & 1.04033 & 1.44000 & $117.897^{+0.157}_{-0.091}$ \\
        10.00 & 1.04033 & 1.50000 & $106.459^{+0.207}_{-0.004}$ \\
        12.00 & 1.04033 & 1.80000 & $73.095^{+1.358}_{-0.006}$ \\
        14.00 & 1.04033 & 2.10000 & $56.489^{+1.242}_{-0.006}$ \\
        16.00 & 1.04033 & 2.40000 & $45.982^{+1.530}_{-0.008}$ \\
        \hline 
        6.00 & 1.05548 & 1.05000 & $\cdots$ \\
        7.00 & 1.05548 & 1.22500 & $354.118^{+0.307}_{-0.633}$ \\
        7.20 & 1.05548 & 1.26000 & $248.950^{+1.203}_{-0.005}$ \\
        7.40 & 1.05548 & 1.29500 & $206.064^{+1.479}_{-0.006}$ \\
        7.60 & 1.05548 & 1.33000 & $179.815^{+0.484}_{-0.006}$ \\
        8.00 & 1.05548 & 1.40000 & $146.516^{+0.354}_{-0.096}$ \\
        9.00 & 1.05548 & 1.57500 & $104.166^{+0.361}_{-0.006}$ \\
        10.00 & 1.05548 & 1.75000 & $82.275^{+0.924}_{-0.007}$ \\
        11.00 & 1.05548 & 1.92500 & $68.351^{+1.485}_{-0.007}$ \\
    \hline    
    \hline
    \end{tabular}
\end{table}

\subsection{Spinning Scattering Simulations}
\begin{figure}[t]
	\includegraphics[width=0.48\textwidth]{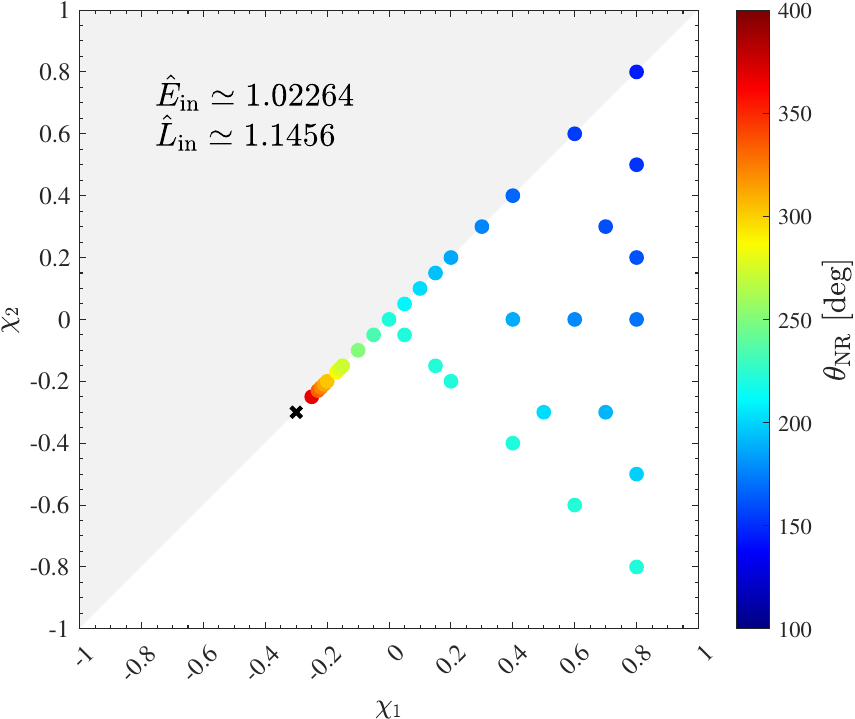}
	\caption{
 \label{fig:chiNR_spin}
NR scattering angles for equal-mass simulations with $\hat{E}_{\rm in} \simeq 1.02264$ and $\hat{L}_{\rm in} \simeq 1.1456$ against individual spin variables.
This figure shows (see the color code) that the angles essentially depend only on the sum of the spins. 
}
\end{figure}

We present here the results of 35 NR simulations of hyperbolic encounters of (equal-mass) spinning BBHs. 
As far as we know, it is the first time unequal-spin simulations have been published.
In particular, we compute a suite of simulations at fixed energy and (canonical) orbital angular momentum and various spin magnitudes. The main parameters of these simulations are reported in Table~\ref{tab:NRspin}, while the scattering angles, as a function of the rescaled spin variables $\chi_i$ are shown in Fig.~\ref{fig:chiNR_spin}.

When unequal spins are considered, the system ceases to be symmetric, inducing a non-zero recoil on the system center-of-mass. The effect of recoil on the scattering has been investigated in Ref.~\cite{Bini:2021gat}. As a consequence of Eq.~(3.33) 
there, the scattering angles measured using the individual trajectories of the two bodies will be different. 
In the following, we will denote by $\theta_{\rm NR}$ the {\it relative} scattering angle, which, for equal-mass binaries, is simply given (to linear order in the recoil) by
\begin{equation}
    \theta_{\rm NR} = \frac{\theta_1 + \theta_2}{2} \, .
\end{equation}

\begin{table}[htbp]
    \centering
        \caption{
        \label{tab:NRspin}
        NR equal-mass, spinning simulations with (approximately) fixed incoming energy $\hat{E}_{\rm in}^{\rm{ADM}} \simeq 1.02264$ and fixed orbital angular momentum $\hat{L}_{\rm in}^{\rm{ADM}} = 1.14560$. 
        The columns, from left to right, represent: the individual spins;
the incoming (rescaled) ADM energy; the incoming (rescaled) orbital ADM angular momentum; 
the scattering angle of the two individual objects (replaced by dots when the objects plunge); and the 
average scattering angle $\theta_{\rm{NR}}$ with its two-sided uncertainty. 
}
    \renewcommand*{\arraystretch}{1.5}
    \setlength{\tabcolsep}{0.1cm}
    \begin{tabular}{c c c c c c c}
    \hline
    \hline
    $\chi_1$ & $\chi_2$ & $\hat{E}_{\rm in}^{\rm{ADM}}$ & $\theta_{1} \left[\rm deg\right]$ &  $\theta_{2} \left[\rm deg\right]$ & $\theta_{\rm{NR}} \left[\rm deg\right]$   \\
    \hline
    -0.30 & -0.30 & 1.02269 & $\cdots$ & $\cdots$ & $\cdots$ \\
    -0.25 & -0.25 & 1.02268 & 367.529 & 367.562 & $\phantom{\ast} 367.545^{+0.000}_{-4.840} \ast$  \\
    -0.23 & -0.23 & 1.02267 & 334.344 & 334.346 & $334.345^{+0.084}_{-1.573}$ \\
    -0.22 & -0.22 & 1.02267 & 322.693 & 322.693 & $322.693^{+0.099}_{-1.004}$ \\
    -0.21 & -0.21 & 1.02267 & 312.795 & 312.795 & $312.795^{+0.187}_{-0.364}$ \\
    -0.20 & -0.20 & 1.02266 & 303.910 & 303.858 & $303.884^{+0.222}_{-0.466}$ \\
    -0.17 & -0.17 & 1.02266 & 286.603 & 286.604 & $286.603^{+0.154}_{-0.010}$ \\
    -0.16 & -0.16 & 1.02266 & 277.848 & 277.850 & $277.849^{+0.230}_{-0.003}$ \\
    -0.15 & -0.15 & 1.02265 & 272.602 & 272.603 & $272.603^{+0.260}_{-0.003}$ \\
    -0.10 & -0.10 & 1.02265 & 251.027 & 251.029 & $251.028^{+0.559}_{-0.003}$ \\
    -0.05 & -0.05 & 1.02264 & 234.747 & 234.389 & $234.568^{+0.845}_{-0.003}$ \\
    \phantom{-}0.00 & \phantom{-}0.00 & 1.02264 & 221.822 & 221.823 & $221.823^{+0.762}_{-0.002}$ \\
    \phantom{-}0.05 & \phantom{-}0.05 & 1.02264 & 211.195 & 211.195 & $211.195^{+0.610}_{-0.002}$ \\
    \phantom{-}0.05 & -0.05 & 1.02264 & 221.950 & 221.782 & $221.866^{+0.643}_{-0.002}$ \\
    \phantom{-}0.10 & \phantom{-}0.10 & 1.02265 & 202.763 & 202.453 & $202.608^{+0.388}_{-0.002}$ \\
    \phantom{-}0.15 & \phantom{-}0.15 & 1.02265 & 194.542 & 194.542 & $194.542^{+0.183}_{-0.001}$ \\
    \phantom{-}0.15 & -0.15 & 1.02265 & 222.141 & 221.632 & $221.887^{+0.637}_{-0.002}$ \\
    \phantom{-}0.20 & -0.20 & 1.02266 & 222.148 & 221.490 & $221.819^{+0.863}_{-0.003}$ \\
    \phantom{-}0.20 & \phantom{-}0.20 & 1.02266 & 187.839 & 187.838 & $187.838^{+0.020}_{-0.141}$ \\
    \phantom{-}0.30 & \phantom{-}0.30 & 1.02269 & 176.588 & 176.585 & $176.586^{+0.001}_{-0.653}$ \\
    \phantom{-}0.40 & -0.40 & 1.02274 & 222.488 & 221.206 & $221.847^{+0.849}_{-0.003}$ \\
    \phantom{-}0.40 & \phantom{-}0.00 & 1.02269 & 188.456 & 187.819 & $188.138^{+0.008}_{-0.132}$ \\
    \phantom{-}0.40 & \phantom{-}0.40 & 1.02274 & 167.545 & 167.544 & $167.545^{+0.002}_{-0.947}$ \\
    \phantom{-}0.50 & -0.30 & 1.02275 & 203.759 & 202.484 & $203.121^{+0.477}_{-0.002}$ \\
    \phantom{-}0.60 & -0.60 & 1.02288 & 223.008 & 221.153 & $222.080^{+0.808}_{-0.003}$ \\
    \phantom{-}0.60 & \phantom{-}0.00 & 1.02276 & 178.087 & 177.170 & $177.629^{+0.001}_{-0.645}$ \\
    \phantom{-}0.60 & \phantom{-}0.60 & 1.02288 & 154.139 & 154.139 & $154.139^{+0.005}_{-1.443}$ \\
    \phantom{-}0.70 & -0.30 & 1.02284 & 191.175 & 189.640 & $190.407^{+0.013}_{-0.164}$ \\
    \phantom{-}0.70 & \phantom{-}0.30 & 1.02284 & 161.186 & 160.685 & $160.935^{+0.004}_{-1.274}$ \\
    \phantom{-}0.80 & -0.80 & 1.02309 & 222.855 & 220.504 & $221.679^{+0.489}_{-0.002}$ \\
    \phantom{-}0.80 & -0.50 & 1.02294 & 199.960 & 198.026 & $198.993^{+0.237}_{-0.001}$ \\
    \phantom{-}0.80 & \phantom{-}0.00 & 1.02287 & 170.873 & 169.914 & $170.394^{+0.003}_{-1.026}$ \\
    \phantom{-}0.80 & \phantom{-}0.20 & 1.02288 & 162.421 & 161.716 & $162.069^{+0.005}_{-1.308}$ \\
    \phantom{-}0.80 & \phantom{-}0.50 & 1.02295 & 152.464 & 152.143 & $152.303^{+0.006}_{-1.640}$ \\
    \phantom{-}0.80 & \phantom{-}0.80 & 1.02309 & 145.467 & 145.248 & $145.357^{+0.006}_{-1.528}$ \\
    \hline
    \hline
    \end{tabular}
\end{table}

\section{NR-deduced physical observables}
\label{sec:nrinfo}

In this section, we discuss some observables that can be directly extracted from the NR simulations presented  in Sec.~\ref{sec:nrsim} above by using only minimal analytical assumptions.

\subsection{Critical orbital angular momentum $L_0$ in nonspinning scattering}

We define the critical (canonical) angular momentum $L_0$ of nonspinning binaries as the value marking the boundary between scattering and plunging systems. Several works (both  analytical \cite{Eardley:2002re,Yoshino:2002tx,Giddings:2004xy,Amati:2007ak,Damour:2017zjx} and numerical \cite{Pretorius:2007jn,Shibata:2008rq,Sperhake:2009jz,Sperhake:2012me,Damour:2022ybd})
have investigated the value of $L_0$ (and the subtleties
in defining it in view of the importance of radiation
losses).

Here we follow Ref.~\cite{Damour:2022ybd} in
estimating $L_0$ by fitting the sequence of NR scattering angles at fixed energy using a template
expected to capture the singular behavior of $\theta(L) $ as $L \to L_0^+$. Namely, we use (recalling the notation $\hat{L} \equiv L/M^2$)
\begin{equation}
	\label{eq:chifit}
	\theta_{\rm NR}^{\rm fit}(\hat{L}) = \frac{\hat{L}}{\hat{L}_0} \ln \left(\frac{1}{1-\hat{L}_0/\hat{L}}\right)\left[\hat{\chi}_{\rm 3PM}(\hat{L};\hat{L}_0) + 2 \frac{a_4}{\hat{L}^4} \right] \, ,
\end{equation}
where $\hat{\chi}_{\rm 3PM}(\hat{L};\hat{L}_0)$ is a third order polynomial in $1/\hat{L}$ defined so as
to ensure a correct PM expansion up to 3PM, and where ${a_4}$ is a 4PM-level fitting parameter (see Sec. VI A of Ref.~\cite{Damour:2022ybd} for details).

For the lower-energy simulations ($\hat{E}_1 = 1.02264$),  we could improve the estimate of $L_0$ computed in Ref.~\cite{Damour:2022ybd} and get
\begin{align}
\label{eq:j0_old}
\hat{L}_0^{\rm fit} (\hat{E}_1) &= 1.0787 \pm 0.00028\,, \nonumber \\
a_4^{\rm fit} (\hat{E}_1 ) &= 6.78 \pm 0.50\,,
\end{align}
The corresponding estimate of $\hat{L}_0^{\rm fit} (\hat{E}_1)$ in \cite{Damour:2022ybd} is 
 $\hat{L}_0^{\rm fit} (\hat{E}_1)= 1.0773$\,.
 
Instead, for the higher-energy ones ($\hat{E}_2 = 1.04033$ and $\hat{E}_3 = 1.05548$, respectively corresponding to center-of-mass velocities $v_{\rm cm,2} \simeq 0.2757$ and $v_{\rm cm,3} \simeq 0.3199$), we got
\begin{align}
\label{eq:j0_new}
\hat{L}_0^{\rm fit} (\hat{E}_2) &= 1.15305 \pm 0.00029\,, \nonumber \\
a_4^{\rm fit} (\hat{E}_2) &= 12.83 \pm 0.81 \,;
\end{align}
and
\begin{align}
\label{eq:j0_newest}
    \hat{L}_0^{\rm fit} (\hat{E}_3 ) &= 1.21431 \pm 0.00036\,,  \nonumber \\
    a_4^{\rm fit} (\hat{E}_3) &= 20.2 \pm 1.8\,.
\end{align}

\begin{figure}[t]
 \includegraphics[width=0.48\textwidth]{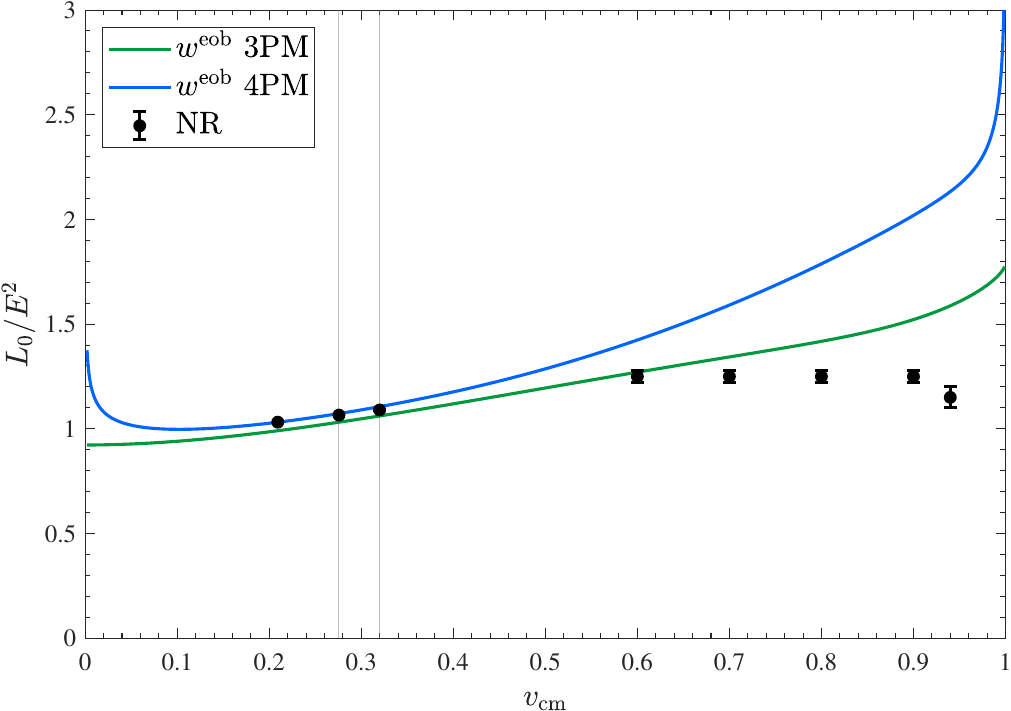}
	\caption{
 \label{fig:j0}
 Comparison between the critical (rescaled) orbital angular momentum $L_0/E^2$ extracted from NR simulations and various analytical predictions.
 The three leftmost NR data points correspond to our fitted estimates [Eqs.~\eqref{eq:j0_old}-\eqref{eq:j0_newest}]. The first point updates the  estimate of Ref.~\cite{Damour:2014afa}, while the other two (highlighted by a vertical grey line) explore a new velocity interval. 
 The five points on the right  correspond to the high-energy simulations of Refs.~\cite{Shibata:2008rq,Sperhake:2009jz}.
 We also display, for comparison, some EOB-PM predictions.}
\end{figure}

In Fig.~\ref{fig:j0} we display these numerical estimates against $v_{\rm cm}$. Note that the two new values $\hat{L}_0^{\rm fit} (\hat{E}_2)$ and 
 $\hat{L}_0^{\rm fit} (\hat{E}_3)$
 fill a gap in the previous knowledge of the critical $\hat{L}_0$ for intermediate energies.
Figure~\ref{fig:j0} also displays the critical angular momentum estimates obtained through the high-velocity simulations of Refs.~\cite{Shibata:2008rq,Sperhake:2009jz}.
We compare these to analytical estimates based on
using the EOB transcription of PM scattering results ($w^{eob}$ nPM). 
For more details on these, see Sec. IV of Ref.~\cite{Damour:2022ybd}.

We can also compare the fitted values  $a_4^{\rm fit} (\hat{E}_n)$, $n=1,2,3$, or more precisely the corresponding 4PM scattering coefficient, $\theta_4^{\rm fit}(\hat{E}_n)$, to the recently determined (radiation-reacted) analytical value of the function $\theta_4(\hat{E})$ \cite{Bern:2021yeh,Dlapa:2021npj,Dlapa:2022lmu,Bini:2022enm}. We find $\theta_4^{\rm fit}(\hat{E}_1) \simeq 60.66$
[to be compared to $\theta_4^{\rm analyt}(\hat{E}_1)= 58.16$], as well as $\theta_4^{\rm fit}(\hat{E}_2) \simeq 84.93$
[to be compared to $\theta_4^{\rm analyt}(\hat{E}_2)= 77.07$] and $\theta_4^{\rm fit}(\hat{E}_3) \simeq 109.23$
[to be compared to $\theta_4^{\rm analyt}(\hat{E}_3)= 94.34$].

When considering spinning systems, the critical curve $L_0/E^2=F(\hat{E})$ becomes a critical surface $L_0/E^2=F(\hat{E}, \chi_1,\chi_2)$. In the following subsection we show how to obtain one point on this critical surface.

\subsection{Spin dependence of the scattering angle}
\label{sec:NR_spin}

We now focus on the spinning simulations and try to determine from our NR results (with minimal analytical assumptions) the spin dependence of the scattering angle at fixed initial energy and angular momentum.
The color coding of Fig.~\ref{fig:chiNR_spin} makes it apparent that the scattering angle mainly depends
on the sum of the spins, i.e. is nearly constant
along the lines $\chi_1+\chi_2=cst$. In particular, the simulations with opposite spins ($\chi_1 = -\chi_2$, i.e. the second diagonal in Fig.~\ref{fig:chiNR_spin}) all lead to scattering angles $\theta_{\rm NR}$ very close (within the percent level) to the nonspinning one at the center.
This numerical result corresponds to the well-known analytical fact (see, e.g., \cite{Damour:2001tu}) that
the leading-PN value of the linear-in-spin
interaction potential (spin-orbit) depends on the effective spin
${\bf S}_{\rm eff}=[1 + 3 m_2/(4 m_1)] {\bf S}_1 + [1 + 3 m_1/(4 m_2)] {\bf S}_2$, while the leading-PN value of the spin-spin interaction potential depends on ${\bf S}_{0}=(1 +  m_2/m_1) {\bf S}_1 + (1 +  m_1/m_2) {\bf S}_2$. In the equal-mass case considered here, this means that the two leading-order spin-dependent interactions only depend on the total spin ${\bf S}_{\rm tot}= {\bf S}_1 +  {\bf S}_2$,
and therefore only on $\chi_1+\chi_2$ in the spin-aligned case.
We have found that this property remains true to a very good approximation, much beyond the leading-order PN approximation. As we will describe below, we have derived a high-PM-accuracy Hamiltonian incorporating the state of the art analytical knowledge of PM gravity, and we found (as displayed in Table~\ref{tab:chiPM_app_spin} in Appendix~\ref{sec:PMtables}) that the averaged scattering angle of two (equal-mass) spinning BHs is analytically predicted to depend essentially only on the average spin $\chi_+ \equiv \frac12 (\chi_1+\chi_2)$. 

Let us also recall that the spin-orbit
interaction term
\begin{equation}
  H_{\rm SO} \simeq +\frac{2 G}{c^3 r^3} {\bf L} \cdot {\bf S}_{\rm eff}
\end{equation}
is {\it repulsive} (respectively, attractive) for spins parallel (respectively, antiparallel) to the orbital angular momentum ${\bf L}$. This fact has a strong
effect on the last stable orbits of quasi-circular spinning binaries (e.g., \cite{Damour:2001tu,Campanelli:2006uy}). In the present, scattering situation, it implies that the leading-order spin-dependent contribution to the scattering angle is negative (positive) when ${\bf L} \cdot {\bf S}_{\rm eff} >0$ (${\bf L} \cdot {\bf S}_{\rm eff} <0$), see e.g.~\cite{Bini:2017wfr}.
This leading-order PN analytical prediction holds also true for the predictions from the PM-accurate EOB Hamiltonian derived below, as can be seen  in the results displayed in Table~\ref{tab:chiPM_spin} in Appendix~\ref{sec:PMtables}.

Our numerical results (displayed in Fig.~\ref{fig:chiNR_spin}) agree remarkably well with the analytical expectation that the (average) scattering angle depends essentially on the average spin, 
\begin{equation}
  \chi_{+} \equiv \frac12 ( \chi_1 +  \chi_2)\,,   
\end{equation}
with only a weak dependence on the antisymmetric spin combination
\begin{equation}
  \chi_{-} \equiv \frac12 ( \chi_1 -  \chi_2)\,.   
\end{equation}
In particular, the numerical results displayed in Table II show that, for most of the (half) spin differences explored by our simulations, the dependence of $\theta_{\rm NR} \equiv \frac12 (\theta_1+\theta_2)$ on $\chi_{-}$ stays within the numerical uncertainty with which we could determine $\theta_{\rm NR}$. For instance, even when $ \chi_{-} = 0.80$, Table II reports that the difference between $\theta_{\rm NR}(\chi_1=0,\chi_2=0)= 221.823^{+0.762}_{-0.002}$ and $\theta_{\rm NR}(\chi_1=0.80,\chi_2=-0.80 =221.679^{+0.489}_{-0.002}$ is only $0.144 \pm 0.454$. In view of this finding, the present limited precision of our  numerical results does not allow us to meaningfully extract  any reliable estimate of the dependence of the function 
$\theta_{\rm NR}(\chi_+,\chi_-)$ on $\chi_-$.
[We note in passing that, for the equal-mass systems we are considering, the relative (i.e. average) scattering angle
$\theta_{\rm rel}=\frac12 (\theta_1+\theta_2)$  is a symmetric function of $\chi_1$ and $\chi_2$, and therefore an even function of $\chi_-$.]

On the other hand, if we focus (for maximum accuracy) on the equal-spin  subset of our numerical results ($\chi_1=\chi_2$, so that $\chi_+= \chi_1=\chi_2$ and $\chi_-=0$), we can fit our results to various possible analytical templates so as to extract a numerical estimate of the univariate function $\theta_{\rm NR, eq}(\chi_+)= \theta_{\rm NR}(\chi_+,\chi_-=0)$.
Removing the $(\chi_1=\chi_2=-0.17)$ data point (which stood out
as an outlier compared to its neighbouring data points
in all our fits; see also Fig.~\ref{fig:chiPM_spin}),
we found that we could fit within numerical errors\footnote{Note that we are neglecting here the additional source of error linked to the fact that the initial energies of our simulations varies by $\pm O(10^{-3})$ from the averaged energy $E_{\rm in}^{\rm av}= 1.0227$.} the remaining eighteen equal-spin data points to the following simple five-parameter template: 
\begin{align}
\label{eq:spinfit}
    \theta_{\rm NR, eq}^{\rm fit}(\chi_+) =&\, \frac{1 - {\rm ln}\left(\chi_+ - \chi_+^{\rm crit}\right)}{1 - {\rm ln} \left(- \chi_+^{\rm crit}\right)}\nonumber \\    
    &\times \left(\theta_0 + \theta_1 \, \chi_+ + \theta_2 \, \chi_+^2 + \theta_3 \, \chi_+^3\right) \, .
\end{align}    
Note that this template incorporates a logarithmic singularity at some critical spin $\chi_+^{\rm crit}$. 
This singularity  models the image on the $\chi_+$ axis of the logarithmic singularity generically expected when crossing the border between scattering and coalescence \cite{Damour:2022ybd}. 
We find the best-fit values (when measuring angles in {\it radians}) to be
\begin{align}
\label{eq:chicrit}
  \theta_0  &= 3.8748 \pm 0.0017 \, , \nonumber \\
  \theta_1  &= 1.881 \pm 0.018 \, , \nonumber \\
  \theta_2  &= 0.43 \pm 0.10 \, , \nonumber \\
  \theta_3  &= 1.07 \pm 0.13 \, , \nonumber \\
  \chi_+^{\rm crit}  &= -0.2932 \pm 0.0013 \, ,
\end{align}
leading to a satisfactory reduced chi-squared: $\chi^2 / (18 - 5) \simeq 0.63$.

The successful fit of our numerical results to the template~\eqref{eq:spinfit} allows us to meaningfully extract from our NR results an estimate of the {\it critical} value of $\chi_+$ (at least in the equal-spin case) leading to immediate plunge rather than scattering. Indeed, the value $\chi_+^{\rm crit} =  -0.2932 \pm 0.0013$  formally corresponds to an infinite scattering angle.
This result completes the critical values of $L_0$ displayed in Fig.~\ref{fig:j0} by yielding the numerical estimate of one point on the surface $L_0^{\rm crit}/E^2=F(\hat E, \chi_1,\chi_2)$ describing the critical initial data for spinning BBHs leading to immediate coalescence, namely the point
\begin{equation}
L_0^{\rm crit}/E^2 = 1.0953, \, \hat{E} = 1.02264, \, \chi_1=\chi_2=-0.2932 \,.
\end{equation}

The computations above have been based on the relative scattering angle $\theta_{\rm rel}=\frac12 (\theta_1+\theta_2)$. We leave to future work a NR-based study
of the dissymmetry between $\theta_1$ and $\theta_2$ 
linked to asymmetric-spin recoil effects.

\section{PM-based analytical predictions}
\label{sec:comp}

In this section, we compare the NR data presented in Sec.~\ref{sec:nrsim} to analytical PM results for the scattering angle of a BBH system on hyperbolic orbits.
For nonspinning BHs, these terms are known up to 4PM, including radiation-reaction effects~\cite{Bern:2021yeh,Bern:2021dqo,Dlapa:2021vgp,Manohar:2022dea,Dlapa:2022lmu,Bini:2022enm}.
The PM formalism has also been extended to include spinning objects, with PM accuracies depending on the order in spin ~\cite{Bini:2017xzy,Bini:2018ywr,Vines:2017hyw,Vines:2018gqi,Guevara:2019fsj,Kalin:2019inp,Kosmopoulos:2021zoq,Chen:2021kxt,Aoude:2022thd,Jakobsen:2022fcj,Bern:2020buy,Bern:2022kto,FebresCordero:2022jts,Alessio:2022kwv,Alessio:2023kgf}.

\subsection{Post-Minkowskian expanded scattering angles}

In the nonspinning limit, the PM approximation (which is a power series in the gravitational constant $G$) translates to a power series in $1/\ell$, where $\ell$ denotes the rescaled orbital angular momentum,
\begin{equation}
    \ell = \frac{L}{G \mu M}\,.
\end{equation}
The total (nonspinning) scattering angle up to $n$PM order (included) can then be written as
\begin{equation}
	\theta_{n{\rm PM}}^{\rm orb}(\gamma,\ell) \equiv \sum_{k = 1}^{n} 2 \frac{\theta_k^{\rm orb}(\gamma)}{\ell^k}\,,
 \end{equation}
where $\gamma$ denotes the Lorentz factor between the two incoming worldlines. We recall that $\gamma$ is equal to the rescaled EOB effective energy, $\gamma=E^{\rm eob}_{\rm eff}/\mu$, which is related to the (real) total energy of the system by
\begin{equation}
    \gamma = \frac{E^2-m_1^2-m_2^2}{2 m_1 m_2}\, .
\end{equation}

When considering spinning BHs, the PM expansion is modified because of the spin-induced multipolar structure of BHs. More precisely,
the multipolar structure of Kerr black holes is polynomial in the Kerr parameters $a_i=S_i/m_i$ (with dimension of length), so that each spin-order, $O(S^m)$ involves a $m$-th order polynomial in $a_1$ and $a_2$.
 In the current PM literature on spinning bodies, it is
usual to consider the Kerr parameters (or ``ring radii")
 $a_i$, as expansion parameters independent of the basic PM expansion parameters $Gm_1$, $Gm_2$. In other words, one uses a double expansion in powers of $Gm_i$ (PM expansion),
 and in powers of $a_i$ (spin expansion). This leads to a double counting $(n,m)$, in which the PM order $n$ counts the (sum of the) powers of $Gm_i$, while $m$ counts the (sum of the) powers of $a_i$. However, one must remember that we are interested in the dynamics of spinning BH's for which we have the inequality $a_i \leq G m_i$. Therefore,  a term of order $(Gm)^n a^m$ is of real PM order $n+m$. This issue must be kept in mind, and will come back in our discussion below, but, for compatibility with the literature, we will continue to refer to a term of order $(Gm)^n a^m$ as being formally of $n$-PM order.

The natural dimensionless expansion parameters associated with the PM and spin expansions are $Gm_i/b$ and $a_i/b$ where $b$ denotes the impact parameter. 
We will, instead of $b$, use the (rescaled) angular momentum $\ell \equiv L/(G m_1 m_2)$, with $L= P_{\rm cm} b$. 
We also utilize $S_i= m_i a_i$ as spin variables, so that each spin order will be identified by a corresponding power of $S_i/\ell$. 
One then writes the combined PM, and spin, expansion of the scattering angle as (in the spin-aligned case)
\begin{equation}
	\label{eq:thetaPM}
	\theta_{n{\rm PM}}(\gamma,\ell,S_i) \equiv \sum_{k = 1}^{n} 2 \frac{\theta_k(\gamma,\ell,S_i)}{\ell^k}\,,
\end{equation}
with the $k$PM contribution to the scattering angle, $\theta_k(\gamma,\ell,S_i)$, being further expanded in spin powers, namely 
\begin{equation}
	\label{eq:thetak}
	\theta_{k}(\gamma,\ell,S_i) \equiv \sum_{m \ge 0} \frac{\theta^{\rm S^m}_k(\gamma)}{\ell^m}\,.
\end{equation}
Here, $\theta^{\rm S^m}_k(\gamma)$ denotes a homogeneous polynomial
in the two spins of order $m$. For instance
\begin{equation}
\theta^{\rm S}_k(\gamma)=\theta^{\rm S_1}_k(\gamma) S_1+\theta^{\rm S_2}_k(\gamma) S_2\,,
\end{equation}
\begin{equation}
\theta^{\rm S^2}_k(\gamma)=\theta^{\rm S_1^2}_k(\gamma) S_1^2+ \theta^{\rm S_1 S_2}_k(\gamma) S_1 S_2+\theta^{\rm S_2^2}_k(\gamma) S_2^2\,.
\end{equation}

As several theoretical papers on spinning scattering
use as basic variables covariantly-defined impact parameters, spins and orbital angular momentum (say $L_{\rm cov}$), while we work here with the 
canonically-defined orbital angular momentum 
($L \equiv L_{\rm can}$), we need to recall the connection between $L_{\rm cov}$ and $L \equiv L_{\rm can}$. In the spin-aligned case, it is simply given by
using the relation \cite{Vines:2017hyw}
\begin{equation}
\label{eq:lcan}
 J=L_{\rm can}+m_1 a_1 + m_2 a_2= L_{\rm cov}  + E_1 a_1+  E_2 a_2\,,
\end{equation}
where $a_i=S_i/m_i$ and $E_i=\sqrt{m_i^2+P_{\rm c.m.}^2}$.
In order to obtain the coefficients $\theta_{k}^{\rm S^m}$ of the double PM- and spin-expansion as functions of the canonical orbital angular momentum, we transform $L_{\rm cov}$ into $L_{\rm can}$ using Eq.~\eqref{eq:lcan}, re-expand in powers of spin, and neglect higher-order contributions.

In the  following, when considering spinning systems, we include: (i)  nonspinning contributions up to 4PM (including radiation-reaction effects)~\cite{Bern:2021dqo,Bini:2021gat,Herrmann:2021tct,Manohar:2022dea,Dlapa:2022lmu,Bini:2022wrq};
(ii) quadratic-in-spin terms up to the 3PM order (including radiation-reaction effects)~\cite{Jakobsen:2022fcj};
(iii) cubic and fourth order in spin effects up
to the 2PM order~\cite{Guevara:2018wpp}. 
In addition, as during the development of this work the
conservative~\cite{Jakobsen:2023ndj},
and radiative~\cite{Jakobsen:2023hig}, linear-in-spin contributions have been computed at the 4PM level, we also incorporated these new analytical results. Their impact is
separately discussed below.

\subsection{$w^{eob}$ resummed scattering angles}

In addition to PM-expanded scattering angles, we also compute $w^{eob}$-resummed PM angles, as introduced in Ref.~\cite{Damour:2022ybd} and generalized here to  spinning bodies.

Let us briefly introduce  the EOB formalism (see, e.g.,  Sec.~IIB of Ref.~\cite{Damour:2022ybd} for more details).
The EOB framework~\cite{Buonanno:1998gg,Buonanno:2000ef,Damour:2000we} is a way of mapping the general-relativistic dynamics of two masses $m_1, m_2$ (considered
in the center-of-mass system) onto the effective relativistic dynamics of 
a single body of mass $\mu= m_1 m_2/(m_1+m_2) = \nu M$. The ``real'' center-of-mass Hamiltonian is related to the ``effective'' 
one through
\begin{equation}
	H_{\rm real} = M \sqrt{1+2\nu\left(\frac{H_{\rm eff}}{\mu}-1\right)}\,.
\end{equation}
For scattering motions, the value of the effective energy $E_{\rm eff}=H_{\rm eff}$ is simply related to the
Lorentz factor $\gamma= - u_1 \cdot u_2$ between the two incoming worldlines by $E_{\rm eff}= \mu \gamma$. 
This implies that the total ADM energy of the system is related to $\gamma$ via
\begin{equation}
{\hat E} = \frac{E}{M}= \sqrt{1+ 2\nu(\gamma-1)}\, .
\end{equation}
The effective Hamiltonian $H_{\rm eff}$ is obtained
by solving (in $P_0=-H_{\rm eff}$) some relativistic-like mass-shell condition
of the general form
\begin{equation}
0= \mu^2+ g_{\rm eff}^{\mu \nu} P_\mu P_\nu \, ,
+ Q(R,P_\mu)\,
\end{equation}
where $Q(R,P_\mu)$ gathers terms more than quadratic in $P_\mu$.
 When considering scattering problems, it is convenient to use a mass-shell condition where the effective metric is the Schwarzschild metric, and where $ Q(R,P_\mu)$ is expressed as a function of
 the effective energy $- P_0= \mu \gamma$. Using
 isotropic coordinates, and rescaled variables, $\bar{r}=R/M$, $p_\alpha = P_\alpha/\mu$, the PM-expansion of the EOB mass-shell condition formally takes the form of a non-relativistic energy-conservation law, namely~\cite{Damour:2017zjx,Damour:2019lcq}
\begin{equation}
	\label{eq:p2}
	p_{\bar{r}}^2 + \frac{\ell^2}{\bar{r}^2} =p_\infty^2 + w^{eob}(\bar{r}; \gamma) \,,
\end{equation}
where $p_\infty = \sqrt{\gamma^2-1}$ and where
the Newtonian-looking potential $w^{eob}(\bar{r}; \gamma)$ [more precisely, $-w^{eob}(\bar{r})$] encapsulates the 
(relativistic, energy-dependent) attractive gravitational interaction. 
For nonspinning systems, the PM expansion of the
$w^{eob}$ potential reads
\begin{equation}
\label{eq:w0}
    w_{n{\rm PM}}^{\rm orb} (\bar{r},\gamma) = \sum_{k=1}^n \frac{w_k^{\rm orb}(\gamma)}{\bar{r}^k}\,,
\end{equation}
where each order in $\frac1{\bar{r}}$ corresponds to
the same PM order (e.g., the first term, $\frac{w_1^{\rm orb}(\gamma)}{\bar{r}}$,
with $w_1^{\rm orb}(\gamma)= 2 (2 \gamma^2-1)$,  describes the 1PM gravitational
interaction \cite{Damour:2016gwp}).

For aligned-spin binaries we can generalize Eq.~\eqref{eq:w0} 
simply by considering  a spin-dependent radial potential of the form (up to fourth order in the spins)
\begin{align}
\label{eq:wspin}
    w_{n{\rm PM}}(\bar{r},\gamma,\ell,S_i) &=  w^{\rm orb} (\bar{r},\gamma) \nonumber \\
    &+ \frac{\ell \, w_{n{\rm PM}}^{\rm S} (\bar{r},\gamma)}{\bar{r}^2} + \frac{w_{n{\rm PM}}^{\rm S^2} (\bar{r},\gamma)}{\bar{r}^2} \nonumber \\
    &+ \frac{ \ell \, w_{n{\rm PM}}^{\rm S^3} (\bar{r},\gamma)} {\bar{r}^4} + \frac{w_{n{\rm PM}}^{\rm S^4} (\bar{r},\gamma)}{\bar{r}^4}.
\end{align}
Here each $w_{n{\rm PM}}^{\rm S^m}(\bar{r},\gamma)$ is a polynomial of order $m$ in spins, whose coefficients admit an expansion in powers of $\frac{1}{\bar{r}}$
up to the order $\frac{1}{\bar{r}^n}$ included. For instance, we have (at spin orders $m=1$ and $m=2$)
\begin{equation}
    w_{n{\rm PM}}^{\rm S} (\bar{r},\gamma) = \sum_{k=1}^n \frac{w_k^{\rm S_1}(\gamma) S_1+ w_k^{\rm S_2}(\gamma) S_2}{\bar{r}^k}\,,
\end{equation}
\begin{equation}
    w_{n{\rm PM}}^{\rm S^2} (\bar{r},\gamma) = \sum_{k=1}^n \frac{w_k^{\rm S_1^2}(\gamma) S_1^2+ w_k^{\rm S_1 S_2}(\gamma) S_1 S_2 + w_k^{\rm S_2^2}(\gamma) S_2^2}{\bar{r}^k}\,.
\end{equation}

The energy-dependent $w^*_k(\gamma)$ coefficients entering the EOB potentials are in 1-to-1 correspondence with the $\theta^*_k(\gamma)$ coefficients of the scattering angle expansion.
These relations can be found by expanding, and solving, at each order in $G$ and in spin, the following equation
\begin{align}
    \label{eq:chi_pr}
\pi + &\theta \left(\gamma, \ell, S_i\right) = - 2\int_{\bar{r}_{\rm min}}^{+\infty} d\bar{r}\, \frac{\partial p_{\bar{r}}\left(\bar{r},\gamma, \ell, S_i\right)}{\partial \ell} \nonumber \\
&\quad= \int_{\bar{r}_{\rm min}}^{+\infty} \frac{d\bar{r}}{\bar{r}^2} \, \, \frac{2\ell - w^{\rm S}(\bar{r},\gamma)  - w^{\rm S^3}(\bar{r},\gamma)/\bar{r}^2 }{\sqrt{p_\infty^2 - \ell^2/\bar{r}^2 + w(\bar{r},\gamma,\ell,\chi_i)}}\,.
\end{align}
Here $\bar{r}_{\rm min}$ denotes the turning point radius.
The explicit equations linking the $\theta^*_k(\gamma)$ coefficients and the $w^*_k(\gamma)$ ones are given in Appendix~\ref{sec:wchi_coefs}.
When using the existing PM-expanded results for
the scattering angle,
the relations of Appendix~\ref{sec:wchi_coefs}
allow one to compute the explicit values of the
PM-expansion coefficients $w^*_k(\gamma)$ describing
the spin interactions within the EOB formalism.
These values will be found in the ancillary file of this paper. 
Note that the so-constructed EOB potential is
a function of the incoming energy and angular momentum of the system that incorporates both conservative and radiation-reaction effects.

Following Ref.~\cite{Damour:2022ybd}, we then define spin-dependent $w^{eob}$-resummed scattering angles $\theta_{n{\rm PM}}^w(\gamma,\ell,S_i)$  by substituting  the $n$PM-accurate, spin-dependent, EOB potentials 
 $w_{n{\rm PM}}(\bar{r},\gamma,\ell,S_i)$ 
 in Eq.~\eqref{eq:chi_pr}, without performing
 any expansion (in $G$ or in spin), i.e. 
\begin{equation} 
\label{thetaeob}
\pi + \theta^{w^{eob}}_{n{\rm PM}} \equiv \int_{0}^{\bar{u}_{\rm max}} d\bar{u}\, \frac{2\ell - w_{n{\rm PM}}^{\rm S}  - w_{n{\rm PM}}^{\rm S^3} \, \bar{u}^2 }{\sqrt{p_\infty^2 - \ell^2 \bar{u}^2 + w_{n{\rm PM}}}}\,,
\end{equation}
where we introduced $\bar{u} \equiv 1/\bar{r}$ and $\bar{u}_{\rm max}=1/\bar{r}_{\rm min}$.

\section{Comparing numerical results to analytical predictions}
\label{sec:NRPM}

\subsection{Numerics/analytics comparison for nonspinning simulations}

\begin{figure*}[t!]
	\includegraphics[width=0.49\textwidth]{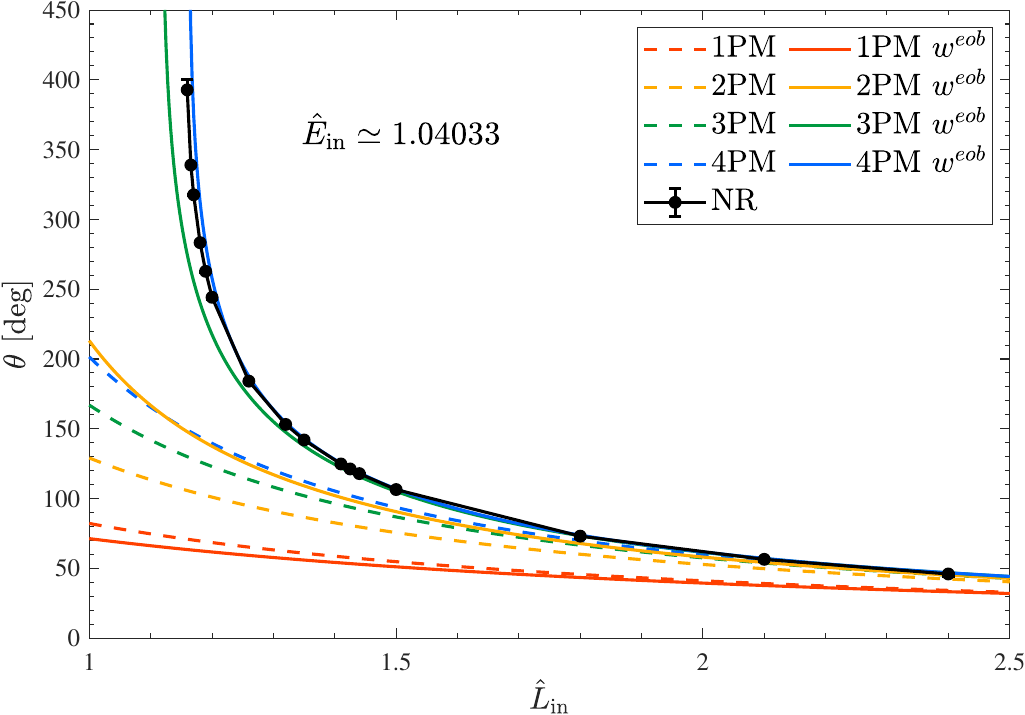}
	\includegraphics[width=0.49\textwidth]{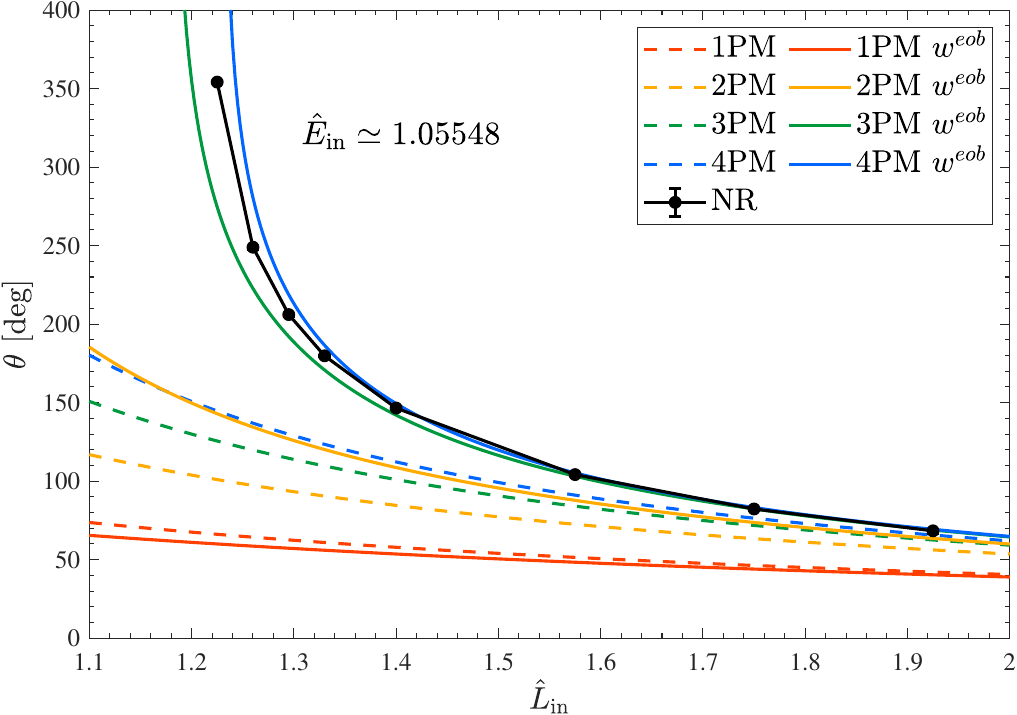}
	\caption{
	Comparison between NR scattering angles of nonspinning BBH systems and analytical PM predictions, both PM-expanded (dashed lines) and $w^{eob}$-resummed (solid lines). Left panel: Intermediate energy $\hat{E}_{\rm in} \simeq 1.04033$. Right panel: Higher energy $\hat{E}_{\rm in} \simeq 1.05548$.
 The $w^{eob}$-resummed 3PM and 4PM results show a remarkable agreement with numerical down to strong-field regimes.
	}
	\label{fig:chiPM_nospin}
\end{figure*}

We start by comparing, in Fig.~\ref{fig:chiPM_nospin}, the results of the nonspinning NR simulations at fixed energy to both the PM-expanded, perturbative,  predictions, Eq.~\eqref{eq:thetaPM}, and their
EOB-resummed versions, defined by Eq.~\eqref{thetaeob}.
This comparison extends the one considered in 
Refs.~\cite{Khalil:2022ylj,Damour:2022ybd}, which 
corresponded to
the lower initial energy $\hat{E}_1$ used in \cite{Damour:2014afa} (see also Appendix~\ref{sec:PMtables}).
The main findings of the latter comparison when considering the PM-expanded perturbative predictions remain true for the new, higher energies considered in this paper, $\hat{E}_2 = 1.04033$ and $\hat{E}_3 = 1.05548$. 
The sequence of PM-expanded results (dashed lines) monotonically approach, in an increasingly accurate
way,  the NR results in the high angular momenta domain (i.e., for relatively large impact parameters fields). However, they perform less and less well
as the angular momentum decreases. Note, in particular, that, for our minimum $\hat{L}$, even the 4PM-accurate prediction differs by a factor $\gtrsim 2$ 
from the NR result.

On the other hand, the sequence of  $w^{eob}$-resummed (PM-based) predicted angles (solid lines) perform much
better than their perturbative counterparts. While the
1PM-based $w^{eob}$-resummed prediction is not better
(and even slightly worse) than its perturbative analog, the 2PM-based $w^{eob}$-resummed fares as well as the 4PM perturbative one, and the 3PM and, especially, 4PM EOB-resummed results exhibit
a remarkable  agreement with NR data for essentially
all orbital angular momenta. It should, however, be noted that the 4PM-based EOB-resummed predicted angles
start to show visible discrepancies with NR results 
for the lowest $L$'s, especially for the highest
energy $\hat{E}_3= 1.05548$. More precisely, $w^{eob}_{\rm 4 PM}$ somewhat overestimates the scattering angle,
while $w^{eob}_{\rm 3 PM}$ underestimates it. We discuss below how these findings can lead to NR-based
ways of improving the knowledge of the exact
$w^{eob}(\bar{r}, \gamma)$ potential.

\subsection{Extracting from NR results the gravitational interaction potential between nonspinning black holes}

\begin{figure*}[t!]
	\includegraphics[width=0.48\textwidth]{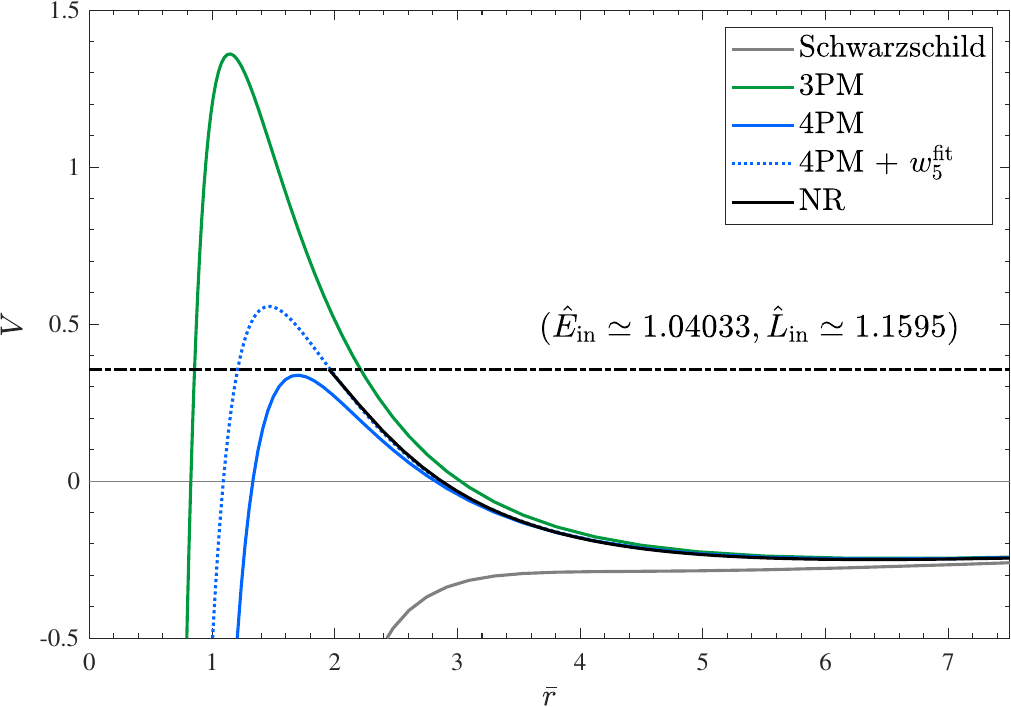}
	\includegraphics[width=0.48\textwidth]{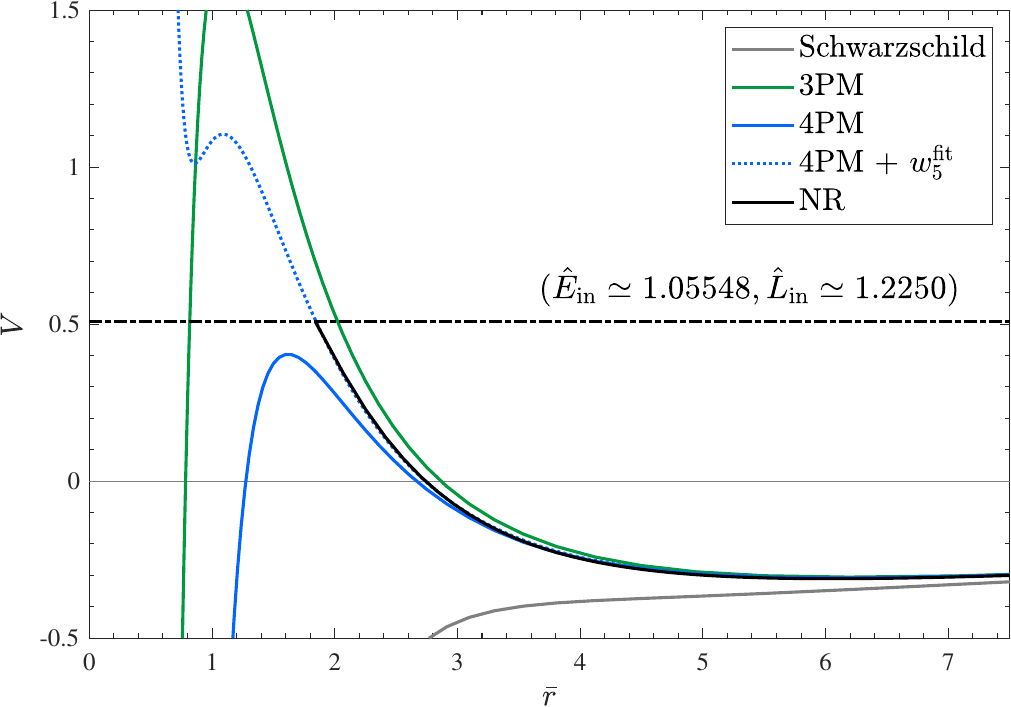}
	\caption{
 Comparison between the gravitational potential $V$ extracted from NR simulations and its EOB-PM equivalent.
 The Schwarzschild potential is plotted as a reference.
Left panel: Intermediate energy $\hat{E}_{\rm in} \simeq 1.04033$. Right panel: Higher energy $\hat{E}_{\rm in} \simeq 1.05548$.
In both cases, the angular momentum is set to the smallest values of angular momentum for which the system scattered.
The horizontal lines mark the values of $p_\infty^2$ for each energy, determining the maximum point up to which it is possible to extract information about the numerical potential. 
}
	\label{fig:wNR}
\end{figure*}

Having computed new suites of numerical simulations with higher initial energies, we can extend the strategy used in Sec. VIC of Ref.~\cite{Damour:2022ybd} and directly extract from
the sequence of NR scattering angles (at a given energy)
the NR radial potential $w_{\rm NR}(\bar{r})$ (in our chosen EOB coordinates). This extraction is done by
iteratively using Firsov's inversion formula~\cite{Landau:1960mec,Kalin:2019rwq} to invert the functional relation $w(\bar{r}, \gamma) \to \theta(\ell,\gamma)$, Eq.~\eqref{eq:chi_pr} (for nonspinning systems), i.e. to deduce the function 
$w_{\rm NR}(\bar{r})$ leading to the function 
$\theta_{\rm NR}(\ell)$, tabulated in Table~\ref{tab:NRnospin} (for each fixed value of $\gamma$). 

The so-extracted NR radial potential for the lower
 value $\gamma_1(E_1) \simeq 1.09188$ has
 been displayed in Fig.~7 of Ref.~\cite{Damour:2022ybd}.
 See Appendix~\ref{sec:PMtables} for an updated
 determination of $w_{\rm NR}(\bar{r}, \gamma_1)$.
We focus here on the determination of the NR 
potentials $w_{\rm NR}(\bar{r}, \gamma_2)$ and
$w_{\rm NR}(\bar{r}, \gamma_3)$ corresponding to the
two higher-energy sequences of simulations reported
in the present work. In addition, we compare the
so-extracted NR gravitational potentials to their
analytical counterparts, as defined by the PM-based
EOB radial potential $ w^{eob}(\bar{r},\gamma)$
defined in Eq.~\eqref{eq:w0} above (for the nonspinning case).

Instead of plotting the various energy-dependent
radial potentials $ w^{*}(\bar{r},\gamma)$, it is
useful to plot the corresponding ``effective radial
potentials", completed by the additional centrifugal
potential $ +\frac{\ell^2}{\bar{r}^2}$. In other words,
we shall compare the (Newtonian-like) effective potentials
\begin{equation}
\label{eq:VEOB}
	V^{*}(\bar{r},\gamma,\ell) \equiv \frac{\ell^2}{\bar{r}^2} - w^{*}(\bar{r},\gamma)\, ,
 \end{equation}
made by adding the (repulsive) centrifugal barrier term $\ell^2 \, \bar{r}^{-2}$ to the (attractive) gravitational radial potential $ - w^{*}(\bar{r},\gamma)$.

The comparison between $V^{\rm NR}(\bar{r},\gamma,\ell)$
and $V^{eob}_{n{\rm PM}}(\bar{r},\gamma,\ell)$ 
(for $n=3$ and $n=4$) is displayed in the
two panels of Fig.~\ref{fig:wNR}. 
The left panel corresponds to the intermediate energy $\gamma_2(E_2) \simeq 1.16456$, and uses the smallest NR
orbital angular momentum leading to scattering, namely $\hat{L} \simeq 1.15950$
(corresponding to $\hat{b}_{\rm NR}=7.73$).
The right panel corresponds to the higher energy $\gamma_3(E_3) \simeq 1.22808$, and uses the smallest NR
orbital angular momentum leading to scattering for that energy, namely $\hat{L} \simeq 1.22500$
(corresponding to $\hat{b}_{\rm NR}=7.00$). 
The NR-extracted
potentials are determined only for (isotropic) radii
larger than about $1.95$ (for $E_2$) and about $1.85$ (for $E_3$).

Note that, for both energies, the NR-extracted potential
is very close to the 4PM-based EOB potential down
to radii  $\bar{r} \simeq 2.8 $. However, in the
interval $ 2 \lesssim \bar{r} \lesssim 2.8 $
$V^{eob}_{4{\rm PM}}(\bar{r})$ starts to exhibit
visible differences with $V^{\rm NR}(\bar{r})$.
For the intermediate energy $E_2$ the peak
of $V^{eob}_{4{\rm PM}}$ lies just below the $p_\infty^2$ line, 
meaning that the EOB-4PM potential predicts a plunge instead of 
a scattering for this angular momentum and this energy. By contrast, the 
(less attractive) EOB-3PM potential still predicts
scattering for this angular momentum and this energy.
The situation is similar for the highest energy, $E_3$.
In this case, the NR potential seems to interpolate
between $V^{eob}_{4{\rm PM}}(\bar{r})$
and $V^{eob}_{3{\rm PM}}(\bar{r})$.

These results suggest that it would be useful to
{\it resum} the EOB-PM potentials\footnote{We recall
that we use here PM-expanded EOB radial
potentials $w(\bar{r}) \sim \frac{1}{\bar{r}}
+ \frac{1}{\bar{r}^2} +  \frac{1}{\bar{r}^3} +\cdots $.}, so as to improve
their performances at small radii. Leaving such an
investigation to future work, we shall content ourselves here by exploring the effect of adding
a formal, additional 5PM-level contribution 
$ + \frac{w_5(\gamma)}{\bar{r}^5}$ to the analytically predicted 4PM EOB potential. 
We can use our determination of  $w^{\rm NR}(\bar{r})$ to extract a
best-fit value for the (energy-dependent)
coefficient $w_5(\gamma)$. Fitting (for our
three energies)
$w^{\rm NR}(\bar{r}, \gamma)$ to $w^{eob}_{4{\rm PM}}(\bar{r})+ \frac{w_5(\gamma)}{\bar{r}^5}$
yields the following results:
\begin{align}    
\label{eq:w5}
    w_5^{\rm fit}(\gamma_1) =& -1.00 \pm 0.17\, , \nonumber \\
    w_5^{\rm fit}(\gamma_2) =& -2.16 \pm 0.13\, , \nonumber \\
    w_5^{\rm fit}(\gamma_3) =& -3.47 \pm 0.23\, .
\end{align}
Here the errors are  indicative estimates obtained by
applying the parametric inversion formulas to the confidence
intervals defined by Eqs~\eqref{eq:j0_old}$-$\eqref{eq:j0_newest}.

The negative sign (present at all energies) reflects the fact that the EOB-4PM potential is slightly too attractive at small radii (while the EOB-3PM potential
is too repulsive for all radii). The strange behaviour of $V^{\rm fit}(\bar{r})=V^{eob}_{4{\rm PM}}(\bar{r})- \frac{w_5^{\rm fit}(\gamma)}{\bar{r}^5}$
 at small radii, for the highest energy (right panel of Fig.~\ref{fig:wNR}) indicates that the addition
 of a (repulsive) 5PM term is only physically
 valid in a small range of radii (say in the range
 $ 2 \lesssim \bar{r} \lesssim 2.8 $). Indeed, when
 one uses a non-resummed potential, 
 $w(\bar{r}) = \frac{w_1}{\bar{r}}
+ \frac{w_2}{\bar{r}^2} + \cdots \frac{w_n}{\bar{r}^n} $, the last term determines the strong-field behavior.
In particular, a negative 5PM term entails a (probably unphysical) repulsive core at very small radii. The use
of suitably resummed versions of the EOB-PM potentials
would probably avoid such undesirable features.

\subsection{Comparing spinning simulations to analytical predictions (without 4PM spin-orbit terms)}

\begin{figure}[t]
 \includegraphics[width=0.48\textwidth]{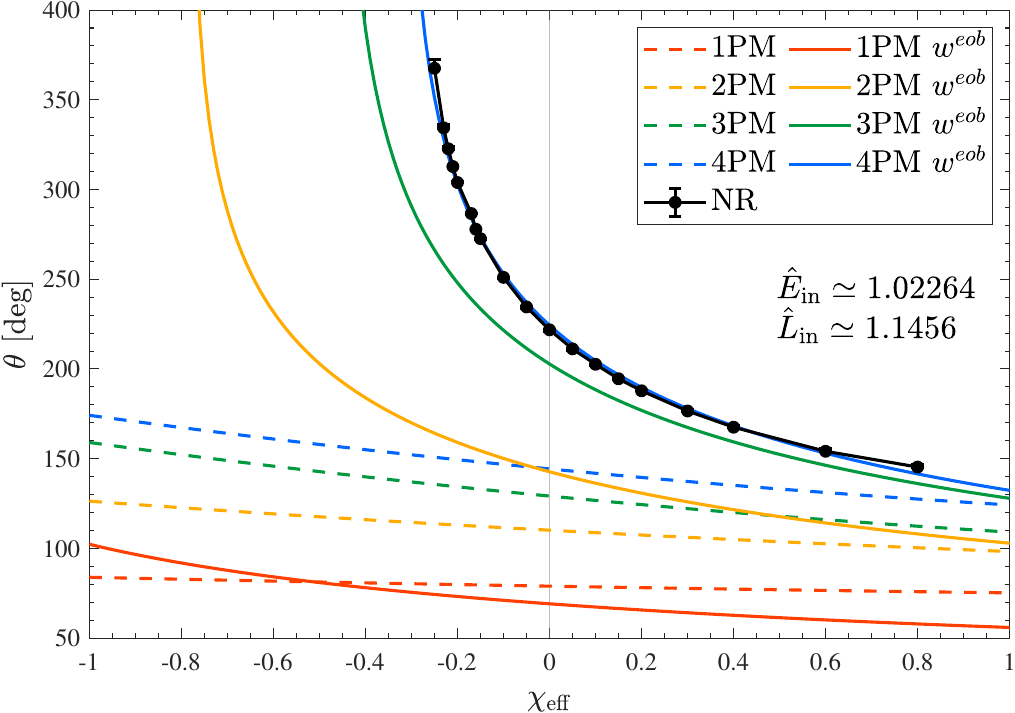}
	\caption{
    \label{fig:chiPM_spin}
	Scattering angle comparison between NR data and both the perturbative, and the EOB-transcribed, PM-based predictions for equal-mass, equal-spin simulations.
 The x-axis represents $\chi_{\rm eff} = \chi_1 = \chi_2$.
 As in Fig.~\ref{fig:chiPM_nospin}, we use dashed lines for the PM-expanded scattering angles, while the solid lines correspond to the $w^{eob}$-resummed ones.
 Similarly to the nonspinning case, the $w^{eob}$-resummed predictions are markedly more accurate, especially the 4PM-informed ones.
	}
\end{figure}

We now shift our attention to the suite of 35 NR simulations which computed
the scattering angles of systems having many different spins, but with fixed initial energy ($\hat{E}_{\rm in} \simeq 1.02264$) and {\it orbital} angular momentum ($\hat{L}_{\rm in} = 1.14560$).  As we have already
seen in Sec.~\ref{sec:nrinfo} that the scattering angle mainly
depends on the average spin $\chi_+$, we actually
focus on the 19 simulations with equal spins (varying between $\chi_1=\chi_2=- 0.30$ and
$\chi_1=\chi_2= + 0.80$) listed in Table~\ref{tab:NRspin}.

In Fig.~\ref{fig:chiPM_spin}, we compare the equal-mass, equal-spin simulations to two types
of PM-informed predictions: (i) standard PM-expanded
angles, of the type of Eq.~\eqref{eq:thetaPM}; 
and (ii) $w^{eob}$-resummed ones [see Eq.~\eqref{thetaeob}]. 
In this section, we do not take into account the recently derived linear-in-spin (spin-orbit) 4PM results of Refs.~\cite{Jakobsen:2023ndj,Jakobsen:2023hig}. 
In other words, we only use at the 4PM level (conservative and radiative) orbital effects.
Since the analytical PM knowledge of the spin-dependent couplings varies with the spin order,
we shall consider, for simplicity, only the following cases: (i) 1PM accuracy in all couplings up to quartic-in-spin terms (labelled as 1PM); 
(ii)
2PM accuracy in all couplings up to $S^4$ (labelled as 2PM);
(iii) 3PM accuracy in the $(S^0,S^1,S^2)$ couplings 
and 2PM accuracy in the $(S^3,S^4)$ ones (labelled as 3PM); 
and (iv) respectively (4PM,3PM,3PM,2PM,2PM) accuracy
in the $(S^0,S^1,S^2,S^3,S^4)$ couplings (labelled as 4PM).

A first conclusion one can draw from Fig.~\ref{fig:chiPM_spin} is that
PM-expanded results are rather unsatisfactory. 
Both the spin-averaged baseline and the spin-dependence are inaccurate, even for the highest
PM accuracy. The agreement becomes, however, more
satisfactory in the high-positive-spin domain, which involves
(because of the repulsive effect of parallel spins)
weaker-field interactions.

By contrast, the $w^{eob}$-resummed angles show, at each PM order, 
a systematically better agreement
with NR data points, especially for the 4PM accuracy which exhibits a remarkable agreement.

\subsection{Inclusion of 4PM spin-orbit terms}
\label{sec:newso}

\begin{figure}[t]
 \includegraphics[width=0.48\textwidth]{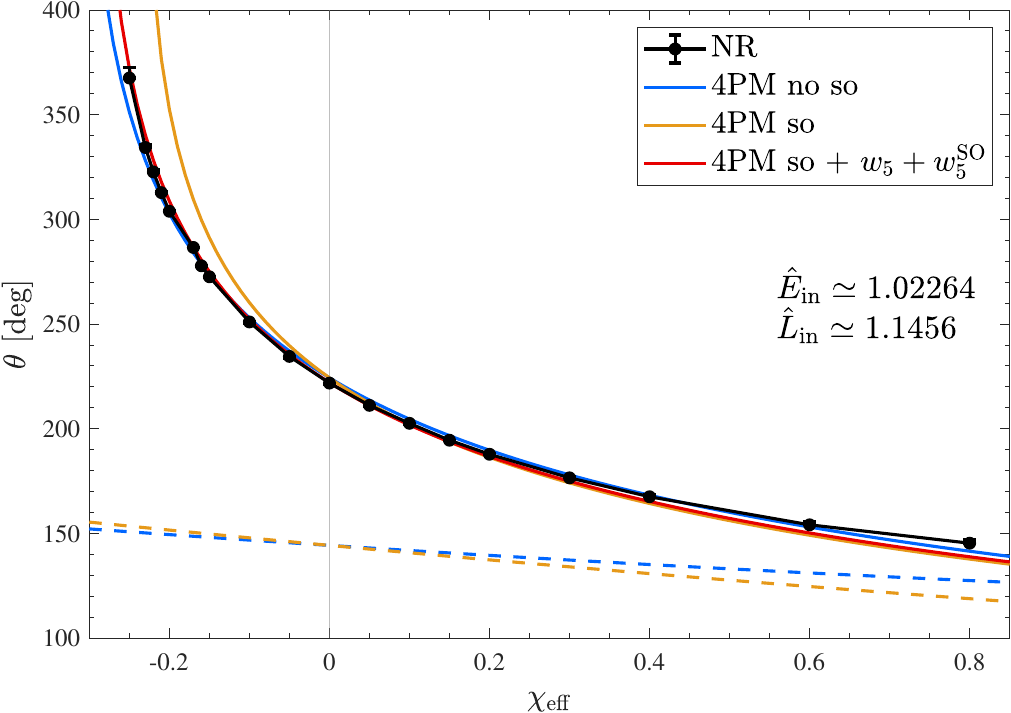}
	\caption{
    \label{fig:chiPM_so} 
Effects due to the inclusion of the 4PM spin-orbit contributions~\cite{Jakobsen:2023ndj,Jakobsen:2023hig} to the scattering angle comparison of Fig.~\ref{fig:chiPM_spin}.
Dashed lines correspond to PM-expanded scattering angles, while solid lines represent $w^{eob}$-resummed predictions.
The blue curves contain spin-dependent terms up to 3PM and nonspinning contributions at 4PM (as in Fig.~\ref{fig:chiPM_spin} above). The orange curves also incorporate (radiation-reacted) spin-orbit effects at 4PM level.
Finally, the red line includes the two additional parameters $w_5$ and $w_5^{\rm SO}$ of Eqs.~\eqref{eq:Dw}-\eqref{eq:w5so}.
	}
\end{figure}

As already mentioned, conservative and radiative 4PM-level linear-in-spin contributions have been computed~\cite{Jakobsen:2023ndj,Jakobsen:2023hig} during
the development of this work. We transformed the latter 
scattering-angle results into additional contributions to the 4PM EOB potential $w_{\rm 4PM}^{\rm S}$. 

In Fig.~\ref{fig:chiPM_so} we compare the effect of incorporating the radiatively corrected 4PM spin-orbit terms 
(``4PM so''), to the 4PM-level angles computed in the previous section which only included orbital effects at 4PM (``4PM no so''). 
For simplicity, we do not separately consider here the conservative contribution to the 4PM spin-orbit terms.

Though the inclusion of 4PM spin-orbit terms has a relatively minor effect in the PM-expanded 
scattering angles (see dashed lines), their inclusion in $w_{\rm 4PM}^{\rm S}$ leads
to much larger differences with respect to the 
corresponding 4PM-level EOB-resummed angles displayed
in Fig.~\ref{fig:chiPM_spin}. 

Looking at Fig.~\ref{fig:chiPM_so}, it is clear that PM-expanded results 
remain rather unsatisfactory, and are significantly
less accurate than $w^{\rm eob}$-resummed ones.
However, the inclusion of 4PM-level spin-orbit effects in $w^{\rm eob}$ 
worsens the EOB-NR agreement that was obtained
when only including 4PM-level orbital coefficients. 

It is possible that the worsening due to the inclusion
of 4PM spin-orbit terms
is a signal pointing out the
need to introduce some resummation of the spin-dependent
 $w_{\rm eob}^{\rm S}$ potential. [We recall that we are
 using here PM-expanded $w_{\rm eob}^{\rm S}$ potentials,
 as indicated in Eq.~\eqref{eq:wspin}.] However, the
 main cause of the worsening effect of the inclusion
of 4PM spin-orbit terms might be the fact, recalled 
in Sec.~\ref{sec:comp} above, that the nomenclature used in current
spin-dependent PM works is physically inappropriate when
dealing with spinning BHs. Indeed, because of the
Kerr BH limit $a_i \leq G m_i$, a term of
formal $n$-PM order including the $m$-th power of spins
is actually of $(n+m)$-PM order. Therefore, the additional
spin-orbit contributions computed in \cite{Jakobsen:2023ndj,Jakobsen:2023hig} are actually
at the physical 5PM order. Similarly, the formally 2PM-level
terms quartic in spins \cite{Guevara:2019fsj} are
actually at the physical 6PM order.

\begin{figure}[t]
 \includegraphics[width=0.48\textwidth]{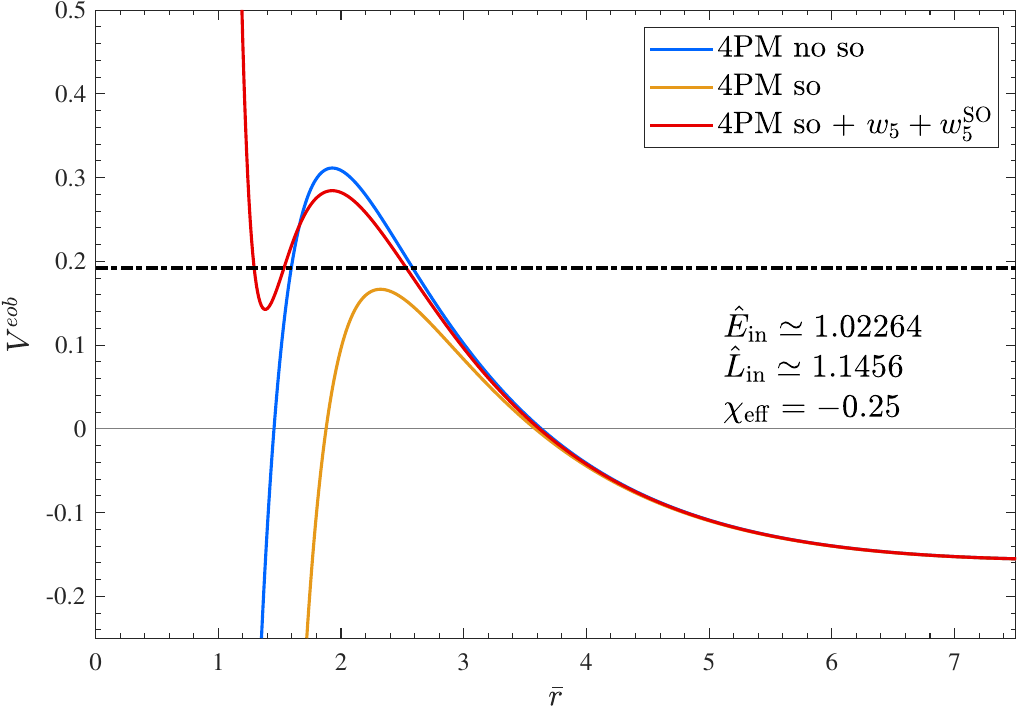}
	\caption{
    \label{fig:VPM_so_calib}
 Comparison of various EOB-PM gravitational potentials $V^{eob}$ in the most negative
 spin case ($\chi_{\rm eff} = -0.25$), i.e. the leftmost NR point in previous figures.
 The displayed analytical potentials correspond to the solid lines in Fig.~\ref{fig:chiPM_so}. 
 The horizontal line marks the ``energy level'' $p_\infty^2$ defining the turning point of the EOB mass-shell 
 condition, Eq.~\eqref{eq:p2}.
	}
\end{figure}

In view of the latter remark, we have explored the effect
of including an additional spin-dependent term 
in $w^{\rm eob}$ 
belonging to the formal 5PM level, but to the physical 6PM level. 
For simplicity, we
only considered a term involving a linear-in-spin contribution of the form
\begin{equation} 
\label{eq:Dw}
 \Delta  w^{\rm eob}  = \frac{w_5}{{\bar r}^5} +  \frac{w_5^{\rm SO} \, \ell \, \chi_{\rm eff}}{{\bar r}^7}\,.
\end{equation}
This additional term contains two dimensionless parameters: 
(i) $w_5$, parametrizing an orbital 5PM-level contribution, and 
(ii) $w_5^{\rm SO}$, parametrizing a spin-orbit contribution at the physical 6PM level\footnote{For simplicity, we do not include here a purely orbital 6PM-level contribution $w_6/\bar{r}^6$.}.

In view of our above finding that the purely orbital
part of the EOB potential is slightly too attractive
and needs to be completed by a $ \frac{w_5(\gamma)}{{\bar r}^5}$
term, where $w_5(\gamma)$ is an energy-dependent negative coefficient,
we fix its value to $w_5^{\rm fit}(\gamma_1)$, Eq.~\eqref{eq:w5}, as the
incoming energy of our spinning simulations corresponds to $\hat{E}_1 \equiv \hat{E}(\gamma_1)$.
Concerning the second,
spin-orbit-related, parameter $w_5^{\rm SO}$ in Eq.~\eqref{eq:Dw},
we determine it by imposing that the corresponding effective Newtonian-like potential 
$V^{eob}$, Eq.~\eqref{eq:VEOB}, predicts immediate coalescence for the critical 
value $\chi_{\rm crit} = -0.2932 \pm 0.0013$, Eq.~\eqref{eq:chicrit}, obtained above by fitting numerical results 
to the simple template of Eq.~\eqref{eq:spinfit}.
This fixes the two parameters entering Eq.~\eqref{eq:Dw} to the approximate values 
\begin{align}
\label{eq:w5so}
w_5 &=-1.00 \pm 0.17\, , \nonumber \\
w_5^{\rm SO} &= +17.3 \pm 0.6 \,.
\end{align}

The corresponding $w^{eob}$-resummed scattering angles computed using the (central) values of 
Eq.~\eqref{eq:w5so} are displayed in Fig.~\ref{fig:chiPM_so}
as a red curve (labelled ``4PM so + $w_5 + w_5^{\rm SO}$").
The addition of the term $\Delta  w^{\rm eob}$, Eq.~\eqref{eq:Dw}, leads to a better visual agreement between NR and EOB, at least for negative spins
(on the large positive spin side, it slightly worsens the NR-EOB difference).

In order to better understand the physical origin of the differences between the various 4PM-level curves displayed in Fig.~\ref{fig:chiPM_so}, we plot in Fig.~\ref{fig:VPM_so_calib} the corresponding Newtonian-like effective potentials $V^{eob}$, Eq.~\eqref{eq:VEOB}. 
Let us recall the meaning of the labels in Fig.~\ref{fig:chiPM_so}: (i) ``4PM no so'' refers to a potential including 
(radiation-reacted) 4PM orbital effects without including any spin-orbit contributions at 4PM;
(ii) ``4PM so'' additionally includes 4PM \textit{radiatively-corrected} spin-orbit terms~\cite{Jakobsen:2023hig};
and finally (iii) ``4PM so + $w_5 + w_5^{\rm SO}$" differs from ``4PM so'' by including the potential $\Delta  w^{\rm eob}$, Eq.~\eqref{eq:Dw}, with the (central) values of 
Eq.~\eqref{eq:w5so}. The dash-dotted horizontal line corresponds to the value of $p_\infty^2$, which defines the turning point of the 
EOB mass-shell condition, Eq.~\eqref{eq:p2}.

The vertical distance between ``4PM no so'' (blue line) and ``4PM so'' (orange line) measures 
the effect of the (radiation-reacted) 4PM spin-orbit contribution.
The inclusion of these terms leads to a more attractive potential. 
Actually, the corresponding potential 
is now below the $p_\infty^2$ ``energy level''  which means that the system is predicted to no longer scatter but to plunge (in agreement in Fig.~\ref{fig:chiPM_so}).

The $\Delta  w^{\rm eob}$ correction makes the potential (red line) more repulsive. 
This leads to a turning
point around $\bar{r} \simeq 2.5$ and to a finite scattering angle. 
This corrected potential also includes a (probably unphysical) repulsive core around $\bar{r} \simeq 1$.

At this stage, the meaning of these results is unclear. 
One will need to explore more values of the incoming energy, of the
incoming angular momentum, and of the spins to assess whether it is indeed
necessary to add (energy-dependent) terms, such as Eq.~\eqref{eq:Dw}, to the current best analytical knowledge of the EOB potential.

\section{Discussion}

In this paper we presented numerical simulations of the scattering of equal-mass BBHs using the \texttt{Einstein Toolkit}~\cite{EinsteinToolkit:2022_11}.

We first computed three sequences of equal-mass, nonspinning BBHs at fixed initial energy and varying angular momentum.
The first sequence, with initial energy $\hat{E}_{\rm in, 1} \simeq 1.02264$, reproduces and extends the results of Ref.~\cite{Damour:2014afa}.
These served as a useful cross-check against previous simulations performed with different numerical codes.
They were also used to improve the estimate of the critical angular momentum $L_0$, marking the boundary between scattering and plunging systems.
We then computed two other sequences of simulations at higher energies, $\hat{E}_{\rm in, 2} \simeq 1.04033$ and $\hat{E}_{\rm in, 3} \simeq 1.05548$, probing stronger-field regimes.
These two series allowed us to compute the critical angular momentum $L_0$ in previously unexplored regions of center-of-mass velocities, $v_{\rm cm,2} \simeq 0.2757$ and $v_{\rm cm,3} \simeq 0.3199$.
This leaves as regions not yet covered by NR simulations of nonspinning BHs the low velocity regime, $v_{\rm cm} \lesssim 0.2$, and the intermediate velocity one, $v_{\rm cm} \simeq 0.3 -0.6$.

We then presented, for the first time, a suite of numerical simulations of equal-mass aligned-spins BHs. 
For these, we fixed both initial energy and angular momentum to be $(\hat{E}_{\rm in},\hat{L}_{\rm in}) \simeq (1.02264,1.1456)$ and varied the individual spins in the range $-0.3 < \chi_i < 0.8$.
We found (as expected both from NR simulations
of coalescing binaries and from analytical predictions) the spin interaction to depend mainly
on the total spin. 
We extracted from numerical data the coefficients
of the linear and quadratic spin dependence of the scattering angle at the considered energy and angular momentum.

We compared our numerical results to PM-based predictions for the scattering angles of equal-mass BHs. We used two types of PM-based analytical predictions: (i) usual, perturbative PM-expanded
results, see Eq.~\eqref{eq:thetaPM}; and (ii) a transcription of PM results 
within the EOB formalism using a (spin-dependent, radiation-reacted)
radial potential in isotropic (EOB) gauge, $w^{eob}(\bar{r}, L, S_1, S_2)$, see Eq.~\eqref{thetaeob}.
We found that PM-expanded scattering angles exhibit
 (both for nonspinning and spinning systems)
 an acceptable agreement with NR data only for large impact parameters, consistently with the PM framework being a weak-field expansion.
On the contrary, a transcription of the PM information into a corresponding (energy-dependent, radiation-reacted, spin-dependent) EOB gravitational potential $w^{eob}(\bar{r}, L, S_1, S_2)$ yields remarkable improvements in the scattering angle comparisons, especially when using
the highest available PM accuracy. See Figs.~\ref{fig:chiPM_nospin} and \ref{fig:chiPM_spin}.
In the case of spinning simulations, we found that the
inclusion of the recently computed 4PM-level spin-orbit contribution
worsened the remarkable agreement (displayed in Fig.~\ref{fig:chiPM_spin}) between NR angles and 
the EOB-resummed 4PM-level ones (without 4PM spin-orbit terms). See Fig.~\ref{fig:chiPM_so}.

Visible discrepancies between NR results and EOB-PM-predicted ones occur only for the few nonspinning cases where
the impact parameter is close to the critical
one leading to plunge rather than scattering 
and, for spinning simulations, when including the 4PM-level spin-orbit contributions.
See Fig.~\ref{fig:wNR}, where we compare (for nonspinning
systems) the EOB-PM potential to a corresponding
potential directly extracted from the NR scattering
data by inverting the relation: potential $\to$ scattering angle. We could, however, improve the
accuracy of the EOB-PM gravitational potential by
adding (an NR-fitted) additional (energy-dependent)
5PM-level contribution $w^{\rm fit}_5(\gamma)/\bar{r}^5$, see Eq.~\eqref{eq:w5}.
Similarly, in the spinning case, we could improve the NR-EOB match obtained when incorporating 
the 4PM-level radiation-reacted spin-orbit terms by including an additional (formally)
5PM-level contribution including a spin-orbit term, Eqs.~\eqref{eq:Dw} and \eqref{eq:w5so}.
The sensitivity of the dynamics to spin-dependent contributions at the 4PM level 
is studied in Fig.~\ref{fig:VPM_so_calib}.

The present work can be extended in several
directions. On the numerical side, a more thorough exploration of the parameter space is called for, notably 
by considering different energies, mass ratios, and spin orientations. In addition, it would be useful
to complete the  computation of the scattering
angles by accurately estimating the emitted
waveform, and by numerically estimating the 
corresponding radiative
losses of energy, linear momentum and angular momentum.

On the analytical side, one should
explore possible resummations of the EOB-PM radial
potential, and generalize it to the non-aligned-spin
case. It would also be interesting to explore other
possible EOB reformulations of perturbative PM results, e.g. using different EOB gauges. Finally, when having in hands NR waveforms, one should compare
them to the recent analytical computations of
$O(G^3)$-accurate, PM-expanded waveforms 
\cite{Brandhuber:2023hhy,Herderschee:2023fxh,Georgoudis:2023lgf}.

Despite the above-mentioned limitations, we hope
that the results presented here will offer a
useful starting point for constructing accurate waveform models for spinning BHs
in eccentric and hyperbolic motions, by combining 
information from NR and PM data within an EOB framework.

\section*{Acknowledgements}

The authors thank R. Russo for useful discussions during the development of this work.
P. R. acknowledges the hospitality and the stimulating environment of the Institut des Hautes Etudes Scientifiques. The present research was partly supported by the {\it ``2021 Balzan Prize for Gravitation: Physical and Astrophysical Aspects"}, awarded to Thibault Damour.
P.~R. acknowledges support by the Fondazione Della Riccia, A.A. 2021/2022. P.~R. is supported by the Italian Minister of University and Research (MUR) via the 
PRIN 2020KB33TP, {\it Multimessenger astronomy in the Einstein Telescope Era (METE)}.
L.M.T. is supported by STFC, the School of Physics and Astronomy at the University of Birmingham, and the Birmingham Institute for Gravitational Wave Astronomy. G.P. is very grateful for support from a Royal Society University Research Fellowship URF{\textbackslash}R1{\textbackslash}221500 and RF{\textbackslash}ERE{\textbackslash}221015, and gratefully acknowledges support from an NVIDIA Academic Hardware Grant. G.P. and P.S. acknowledge support from STFC Grant ST/V005677/1. 
Computations were performed using the University of Birmingham's BlueBEAR HPC service, which provides a High Performance Computing service to the University's research community, the Sulis Tier 2 HPC platform hosted by the Scientific Computing Research Technology Platform at the University of Warwick funded by EPSRC Grant EP/T022108/1 and the HPC Midlands+ consortium, and on the Bondi HPC cluster at the Birmingham Institute for Gravitational Wave Astronomy.

\appendix


\section{Numerical Relativity Systematics}
\label{app:nrerrors}
\subsection{Impact of NR Resolution}
\begin{figure}[th!]
 \includegraphics[width=0.48\textwidth]{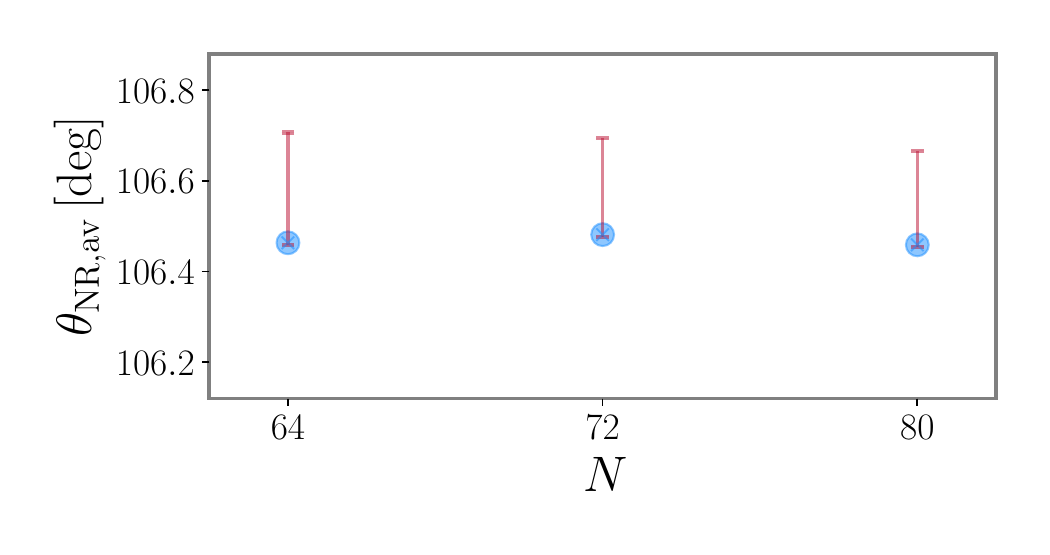}
	\caption{
    \label{fig:nr_error_grid}
	Error on the average scattering angle $\theta_{\rm NR,av}$ derived from the numerical relativity simulations for three different resolutions on the finest grid. Here, $N = 80$ represents the highest resolution, and default choice, whereas $N = 64$ denotes the lowest resolution. For all three resolutions, the upper errors are set by lower-order polynomial fits that overestimate the scattering angle. The lower error bars are set by high-order polynomials that manifestly agree with the preferred order.
	}
\end{figure}

One aspect of our numerical setup that can affect accuracy is the spatial resolution of the simulations. In order to test the robustness of the scattering angle to the NR resolution, we choose a fiducial binary and repeat the scattering simulation at three different resolutions, $N = \left[ 64, 72, 80 \right]$. Here $N$ denotes the number of grid points on the finest level such that the resolution of the smallest box around the puncture is $\sim 2 m_p / N$, with $N = 64$ denoting the lowest resolution and $N = 80$ the highest. Here we take a nonspinning configuration with $P_{\rm ADM} = 0.15$ and $b_{\rm NR} = 10.0$. As shown in Fig.~\ref{fig:nr_error_grid}, we find that the choice of resolution has a minimal impact on the scattering angle, with the errors due to finite NR resolution being subdominant to the errors from the polynomial extrapolation. A summary is given in Table~\ref{tab:nr_res}. 

\begin{table}[th!]
    \caption{
    \label{tab:nr_res}
    A set of equal-mass, nonspinning simulations with an initial momentum $P_{\rm ADM} = 0.175$ and different resolutions. 
    }
    \setlength{\tabcolsep}{0.3cm}
    \begin{tabular}{c c c c c}
    \hline
    \hline   
    $N$ & $\hat{b}_{\rm NR}$ & $\hat{E}_{\rm in}^{\rm{ADM}}$ & $\hat{L}_{\rm in}^{\rm{ADM}}$ & $\theta_{\rm{NR}} \left[\rm deg\right]$ \\
    \hline
    64 & 10.00 & 1.04033 & 1.50000 & $106.463^{+0.243}_{-0.005}$ \\
    72 & 10.00 & 1.04033 & 1.50000 & $106.481^{+0.212}_{-0.005}$ \\
    80 & 10.00 & 1.04033 & 1.50000 & $106.459^{+0.207}_{-0.004}$ \\
    \hline 
    \hline 
    \end{tabular}
\end{table}

\subsection{Impact of Gauge Conditions}
In order to help understand the impact of gauge conditions on our calculation of the scattering angle, we re-run a canonical test simulation, taken to be $b_{\rm NR} = 10.0$ and $\boldsymbol{P}_{\rm ADM} = 0.11456439$, with an alternative lapse profile. The default lapse profile, as noted in earlier, is $\alpha = \psi^{-2}_{\rm BL}$. For the alternative prescription, we adopt the lapse profile proposed in \cite{Tichy:2003zg}, which makes use of an approximate helical Killing vector field to minimise the dynamical evolution of the lapse. Writing the lapse in a pre-collapsed form, we have \cite{Tichy:2003zg} 
\begin{align}
    \label{eq:lapse}
    \tilde{\alpha} &= \frac{1 - \left(\frac{m_1}{2 r_1} + \frac{m_2}{2 r_2} \right)}{1 + \frac{m_1}{2 r_1} + \frac{m_2}{2 r_2}}\, ,
\end{align}
\newline 
where $m_i$ and $r_i$ denote the mass and location of the $i$-th puncture respectively. The lapse is then averaged, $\alpha = (1 + \tilde{\alpha})/2$, to ensure that it lies within a range of $[0, 1]$. The initial shift is taken to be $\beta=0$. 

\begin{figure}[t!]
 \includegraphics[width=0.48\textwidth]{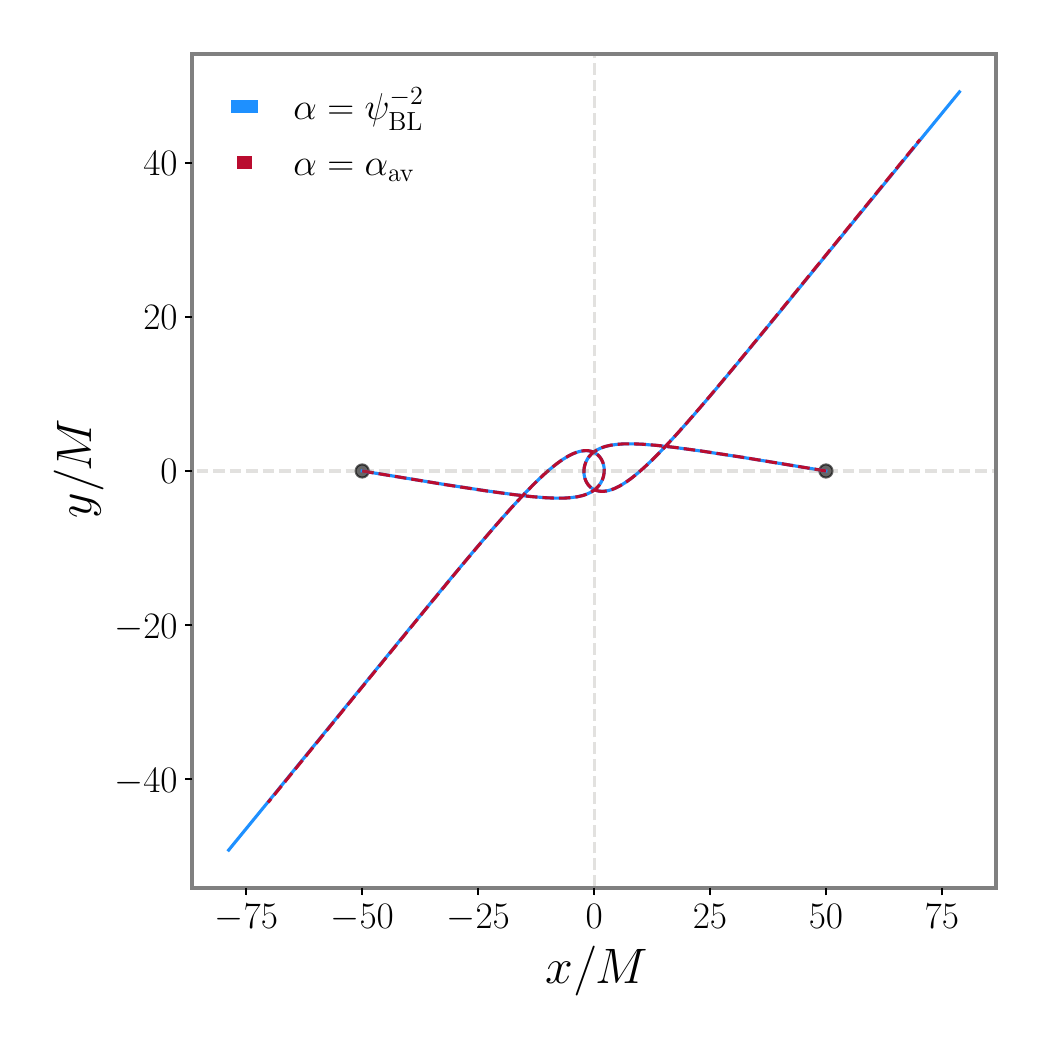}
	\caption{
    \label{fig:nr_error_lapse}
	Trajectories of each BH using two different gauge choices for the initial lapse. The first choice is $\alpha = \psi^{-2}_{\rm BL}$, and the second the averaged lapse defined in Eq.~\eqref{eq:lapse}. The dots denote the starting position of each BH. 
	}
\end{figure}

For this robustness test, we adopt a nonspinning, equal mass binary with $P_{\rm ADM} = 0.11456439$ and $b_{\rm NR}$, as shown in Fig.~\ref{fig:nr_error_lapse}. We find that the choice of lapse again has a minor impact on the calculated scattering angle, finding $\theta_{\rm NR,av} = 221.823^{+0.762}_{-0.002}$ for the default lapse and $\theta_{\rm NR,av} = 222.089^{+0.605}_{-0.002}$ for the alternative lapse. The differences are smaller than the errors arising from the polynomial fitting procedure.  

\section{Mapping between scattering angle and potential coefficients}
\label{sec:wchi_coefs}

We here report the relations between the $w_i$ coefficients entering the radial potential $w_{n{\rm PM}}$, Eq.~\eqref{eq:wspin}, and the $\theta_i$ coefficients constituting the scattering angle $\theta_{n{\rm PM}}$, Eq.~\eqref{eq:thetaPM}.

We start by recalling the equations linking the various energy variables to the $\gamma$ factor of the system:
\begin{align}
    \hat{E}^{\rm eob}_{\rm eff} &\equiv \frac{E^{\rm eob}_{\rm eff}}{\mu} = \gamma\, , \nonumber \\
    \hat{E} &\equiv \,\,\, \frac{E}{M} \,\,\, = \sqrt{1+2\nu\left(\gamma - 1\right)}\, .
\end{align}

The relations between the nonspinning coefficients are known, and read
\begin{align}
    \theta_1^{\rm orb} &= \frac{w_1^{\rm orb}}{2 p_\infty}\, , \nonumber \\
    \theta_2^{\rm orb} &= \frac{\pi}{4} w_2^{\rm orb}\, , \nonumber \\
    \theta_3^{\rm orb} &= \frac{1}{p_\infty^3} \left[p_\infty^4 w_3^{\rm orb} + \frac{1}{2} p_\infty^2 w_2^{\rm orb}w_1^{\rm orb} -\left(w_1^{\rm orb}\right)^3\right]\, , \nonumber \\
    \theta_4^{\rm orb} &= \frac{3 \pi}{8} \left[p_\infty^2 w_4^{\rm orb} +  w_3^{\rm orb} w_1^{\rm orb} + \frac{1}{2} \left(w_2^{\rm orb}\right)^2\right]\, .
\end{align}

In the following, we list the relations containing spin terms relevant to this paper: linear relations up to 4PM, quadratic-in-spin up to 3PM, cubic and quartic-in-spin at 1PM and 2PM.
At linear order in spins we get
\begin{align}
    \theta_1^{\rm S} =& \frac{p_\infty}{2} w_1^{\rm S}\, , \nonumber \\
    \theta_2^{\rm S} =& \frac{\pi}{4} \left(p_\infty^2 w_2^{\rm S} + w_1^{\rm orb} w_1^{\rm S}\right)\, , \nonumber \\
    \theta_3^{\rm S} =& \frac{1}{p_\infty} \Bigg[p_\infty^4 w_3^{\rm S} + \frac{3}{2} p_\infty^2\left(w_2^{\rm orb} w_1^{\rm S}+w_1^{\rm orb} w_2^{\rm S}\right) \nonumber \\
    &+\frac{3}{8} \left(w_1^{\rm orb}\right)^2 w_1^{\rm S}\Bigg]\, , \nonumber \\
    \theta_4^{\rm S} =& \frac{3 \pi}{8} \Bigg[p_\infty^4 w_4^{\rm S} + 2 p_\infty^2\left(w_2^{\rm orb} w_2^{\rm S}+w_3^{\rm orb} w_1^{\rm S}+w_1^{\rm orb} w_3^{\rm S}\right) \nonumber \\
    &+ \left(w_1^{\rm orb}\right)^2 w_2^{\rm S} + 2 w_2^{\rm orb}w_1^{\rm orb}w_1^{\rm S}\Bigg]\, ,
\end{align}
while the quadratic-in-spin terms read
\begin{align}
    \theta_1^{\rm S^2} =& p_\infty w_1^{\rm S^2}\, , \nonumber \\
    \theta_2^{\rm S^2} =& \frac{3\pi}{32} \left[3 p_\infty^2 \left(w_1^{\rm S}\right)^2 + 4 p_\infty^2 w_2^{\rm S^2} + 4 w_1^{\rm orb} w_1^{\rm S^2}\right]\, , \nonumber \\
    \theta_3^{\rm S^2} =& \frac{1}{p_\infty} \Bigg\{\frac{4}{3} p_\infty^4 \left(w_3^{\rm S^2} + 2 w_2^{\rm S} w_1^{\rm S}\right) + \frac{1}{2} \left(w_1^{\rm orb}\right)^2 w_1^{\rm S^2} \nonumber \\
    &+ 2 p_\infty^2 \left[w_2^{\rm orb} w_1^{\rm S^2} + w_1^{\rm orb} \left(w_1^{\rm S}\right)^2 + w_1^{\rm orb} w_2^{\rm S^2}\right]\Bigg\} \, .
\end{align}
The cubic-in-spin and quadratic-in-spin contributions at 1PM order are simply
\begin{align}
    \theta_1^{\rm S^3} =& p_\infty^3 w_1^{\rm S^3}\, , \nonumber \\
    \theta_1^{\rm S^4} =& \frac{4}{3} p_\infty^3 w_1^{\rm S^4}\, ,
    \end{align}
    while the 2PM coefficients can be obtained by
    \begin{align}
    \theta_2^{\rm S^3} =& \frac{3\pi}{8} p_\infty^2 \left(p_\infty^2 w_2^{\rm S^3} + 2 w_1^{\rm orb} w_1^{\rm S^3}
    + 2 w_1^{\rm S} w_1^{\rm S^2} \right)\, , \nonumber \\
    \theta_2^{\rm S^4} =& \frac{15\pi}{64} p_\infty^2 \Bigg[2 p_\infty^2 w_2^{\rm S^4} + 5 p_\infty^2 w_1^{\rm S} w_1^{\rm S^3} \nonumber \\
    &+ 4 w_1^{\rm orb} w_1^{\rm S^4}
    + 2 \left(w_1^{\rm S_2}\right)^2 \Bigg] \, .
\end{align}

The explicit values of the $w_i$ coefficients are given in the ancillary file of this paper.

\section{Additional Post-Minkowskian data points}
\label{sec:PMtables}

In this appendix, we complement the PM results presented in the main paper.

\begin{figure}[t]
	\includegraphics[width=0.49\textwidth]{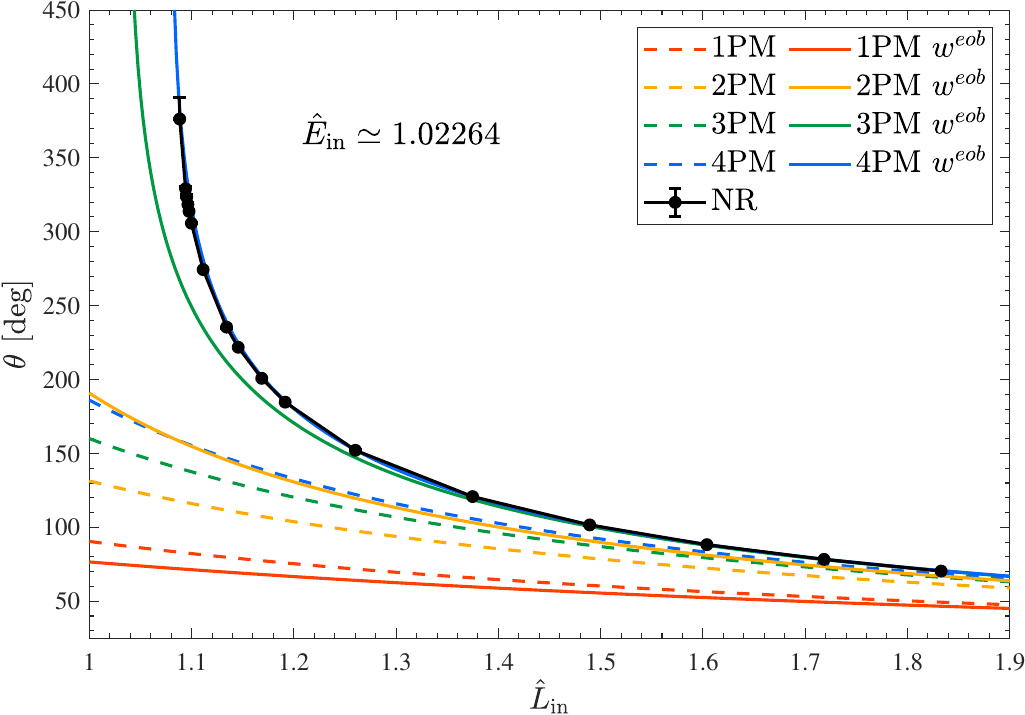}
	\caption{
	\label{fig:thetaPM_app}
 Scattering angles against initial angular momentum (as in Fig.~\ref{fig:chiPM_nospin}) for the lower-energy simulations of equal-mass nonspinning BBHs.
}
\end{figure}

\begin{figure}[t]
	\includegraphics[width=0.48\textwidth]{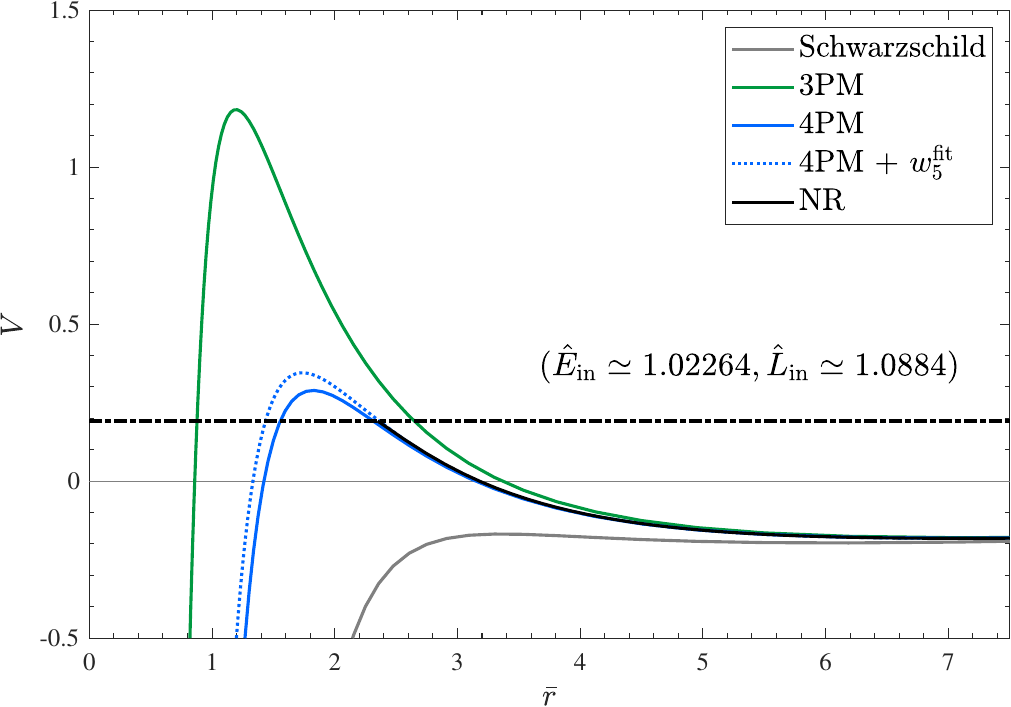}
	\caption{
	\label{fig:wPM_app}
 Gravitational potential $V$ as a function of (center-of-mass, isotropic coordinates) separation (as in Fig.~\ref{fig:wNR}) for the lower energy simulations of equal-mass nonspinning BBHs.
}
\end{figure}

In Figs.~\ref{fig:thetaPM_app} and \ref{fig:wPM_app}, we mimic Figs.~\ref{fig:chiPM_nospin} and \ref{fig:wNR} for the nonspinning systems with lower energy, $\hat{E}_{\rm in,1} \simeq 1.02264$.
These figures confirm the outstanding agreement between EOB-PM predictions and NR results found in Ref.~\cite{Damour:2014afa}.

In Tables~\ref{tab:chiPM_app_nospin} and Tables~\ref{tab:chiPM_app_spin}, we list the scattering angles for the presented NR simulations together with the corresponding best-performing analytical predictions (not considering 4PM spin-orbit terms), $\theta^w_{\rm 3PM}$ and $\theta^w_{\rm 4PM}$. 

\begin{table}[t]
    \caption{
    \label{tab:chiPM_app_nospin}
        NR equal-mass, nonspinning simulations presented in Table~\ref{tab:NRnospin}.
        The last two columns list the $w^{eob}$-resummed scattering angles at 3PM and 4PM as in Figs.~\ref{fig:chiPM_nospin} and \ref{fig:thetaPM_app}. 
        The scattering angles are replaced by dots when the black holes (are predicted to) plunge.
        An asterisk denotes uncertainty about an eventual plunge.
    }
    \setlength{\tabcolsep}{0.15cm}
    \begin{tabular}{c c c c c}
    \hline
    \hline   
    $\hat{E}_{\rm in}^{\rm{ADM}}$ & $\hat{L}_{\rm in}^{\rm{ADM}}$ & $\theta_{\rm{NR}} \left[\rm deg\right]$ & $\theta^w_{\rm 3PM}\left[\rm deg\right]$ & $\theta^w_{\rm 4PM}\left[\rm deg\right]$ \\
    \hline
1.02264 & 1.0769 & $\cdots$ & 288.979 & $\cdots$ \\ 
1.02264 & 1.0884 & $\phantom{\ast} 376.275^{+0.026}_{-14.69} \ast$ & 266.491 & 379.310 \\ 
1.02264 & 1.0941 & $329.057^{+0.003}_{-1.534}$ & 257.257 & 341.135 \\ 
1.02264 & 1.0952 & $323.422^{+0.000}_{-1.914}$ & 255.589 & 335.476 \\ 
1.02264 & 1.0964 & $318.394^{+0.000}_{-1.575}$ & 253.808 & 329.728 \\ 
1.02264 & 1.0975 & $313.764^{+0.000}_{-1.331}$ & 252.209 & 324.803 \\ 
1.02264 & 1.0998 & $305.734^{+0.056}_{-0.694}$ & 248.965 & 315.401 \\ 
1.02264 & 1.1113 & $274.368^{+0.074}_{-0.016}$ & 234.475 & 280.337 \\ 
1.02264 & 1.1342 & $235.447^{+0.912}_{-0.003}$ & 211.830 & 238.507 \\ 
1.02264 & 1.1456 & $221.823^{+0.762}_{-0.002}$ & 202.691 & 224.179 \\ 
1.02264 & 1.1686 & $200.810^{+0.620}_{-0.004}$ & 187.190 & 201.991 \\ 
1.02264 & 1.1915 & $184.684^{+0.221}_{-0.002}$ & 174.595 & 185.411 \\ 
1.02264 & 1.2602 & $152.106^{+0.055}_{-0.446}$ & 147.085 & 152.231 \\ 
1.02264 & 1.3748 & $120.804^{+0.013}_{-0.307}$ & 118.731 & 120.821 \\ 
1.02264 & 1.4893 & $101.616^{+0.059}_{-0.002}$ & 100.654 & 101.704 \\ 
1.02264 & 1.6039 & $88.260^{+0.337}_{-0.002}$ & 87.828 & 88.423 \\ 
1.02264 & 1.7185 & $78.296^{+0.520}_{-0.002}$ & 78.150 & 78.515 \\ 
1.02264 & 1.833 & $70.404^{+0.927}_{-0.003}$ & 70.539 & 70.777 \\ 
\hline
1.04032 & 1.1505 & $\cdots$ & 295.902 & $\cdots$ \\ 
1.04032 & 1.1595 & $\phantom{\ast} 392.815^{+0.006}_{-7.477} \ast$ & 274.866 & $\cdots$ \\ 
1.04032 & 1.1655 & $338.973^{+0.156}_{-0.756}$ & 263.305 & 430.914 \\ 
1.04032 & 1.17 & $317.637^{+0.142}_{-0.444}$ & 255.591 & 374.716 \\ 
1.04032 & 1.1805 & $283.359^{+0.343}_{-0.007}$ & 240.035 & 312.437 \\ 
1.04032 & 1.1895 & $262.825^{+0.749}_{-0.008}$ & 228.777 & 282.302 \\ 
1.04032 & 1.2 & $244.210^{+1.220}_{-0.005}$ & 217.439 & 257.740 \\ 
1.04033 & 1.26 & $184.138^{+0.439}_{-0.004}$ & 173.732 & 187.734 \\ 
1.04033 & 1.32 & $153.119^{+0.226}_{-0.227}$ & 147.539 & 154.603 \\ 
1.04033 & 1.35 & $141.986^{+0.244}_{-0.213}$ & 137.744 & 143.094 \\ 
1.04033 & 1.41 & $124.805^{+0.154}_{-0.238}$ & 122.169 & 125.487 \\ 
1.04033 & 1.425 & $121.233^{+0.180}_{-0.153}$ & 118.900 & 121.880 \\ 
1.04033 & 1.44 & $117.897^{+0.157}_{-0.091}$ & 115.828 & 118.517 \\ 
1.04033 & 1.5 & $106.459^{+0.207}_{-0.004}$ & 105.174 & 107.016 \\ 
1.04033 & 1.8 & $73.095^{+1.358}_{-0.006}$ & 73.374 & 73.826 \\ 
1.04033 & 2.1 & $56.489^{+1.242}_{-0.006}$ & 56.993 & 57.160 \\ 
1.04033 & 2.4 & $45.982^{+1.530}_{-0.008}$ & 46.791 & 46.868 \\ 
\hline
1.05548 & 1.05 & $\cdots$ & $\cdots$ & $\cdots$ \\ 
1.05548 & 1.225 & $354.118^{+0.307}_{-0.633}$ & 274.582 & 391.281 \\ 
1.05548 & 1.26 & $248.950^{+1.203}_{-0.005}$ & 222.675 & 280.445 \\ 
1.05548 & 1.295 & $206.064^{+1.479}_{-0.006}$ & 192.055 & 218.732 \\ 
1.05548 & 1.33 & $179.815^{+0.484}_{-0.006}$ & 170.787 & 186.552 \\ 
1.05548 & 1.4 & $146.516^{+0.354}_{-0.096}$ & 142.049 & 149.418 \\ 
1.05548 & 1.575 & $104.166^{+0.361}_{-0.006}$ & 103.162 & 105.287 \\ 
1.05548 & 1.75 & $82.275^{+0.924}_{-0.007}$ & 82.292 & 83.191 \\ 
1.05548 & 1.925 & $68.351^{+1.485}_{-0.007}$ & 68.882 & 69.339 \\ 
    \hline    
    \hline
    \end{tabular}
\end{table}

\begin{table}[t]
    \centering
        \caption{
        \label{tab:chiPM_app_spin}
        NR equal-mass, spinning simulations listed in Table~\ref{tab:NRspin}.
        All systems were computed with fixed initial ADM energy $\hat{E}_{\rm in}^{\rm{ADM}} \simeq 1.02264$ and orbital ADM angular momentum $\hat{L}_{\rm in}^{\rm{ADM}} \simeq 1.14560$. 
        The last two columns list the $w^{eob}$-resummed scattering angles at 3PM and 4PM as in Fig.~\ref{fig:chiPM_spin}. 
}
    \setlength{\tabcolsep}{0.15cm}
    \begin{tabular}{c c c c c}
    \hline
    \hline
    $\chi_1$ & $\chi_2$ & $\theta_{\rm{NR}} \left[\rm deg\right]$ & $\theta^w_{\rm 3PM}\left[\rm deg\right]$ & $\theta^w_{\rm 4PM}\left[\rm deg\right]$  \\
    \hline
        -0.3 & -0.3 & $\cdots$ & 209.786 & 494.626 \\ 
-0.25 & -0.25 & $\phantom{\ast} 367.545^{+0.000}_{-4.840} \ast$ & 266.889 & 351.203 \\ 
-0.23 & -0.23 & $334.345^{+0.084}_{-1.573}$ & 266.305 & 327.892 \\ 
-0.22 & -0.22 & $322.693^{+0.099}_{-1.004}$ & 258.352 & 318.528 \\ 
-0.21 & -0.21 & $312.795^{+0.187}_{-0.364}$ & 251.257 & 310.189 \\ 
-0.2 & -0.2 & $303.884^{+0.222}_{-0.466}$ & 247.895 & 302.522 \\ 
-0.17 & -0.17 & $286.603^{+0.154}_{-0.010}$ & 238.849 & 283.736 \\ 
-0.16 & -0.16 & $277.849^{+0.230}_{-0.003}$ & 236.080 & 278.413 \\ 
-0.15 & -0.15 & $272.603^{+0.260}_{-0.003}$ & 233.369 & 273.354 \\ 
-0.1 & -0.1 & $251.028^{+0.559}_{-0.003}$ & 221.462 & 252.778 \\ 
-0.05 & -0.05 & $234.568^{+0.845}_{-0.003}$ & 211.364 & 236.915 \\ 
\phantom{-}0.00 & \phantom{-}0.00 & $221.823^{+0.762}_{-0.002}$ & 202.691 & 224.179 \\ 
\phantom{-}0.05 & \phantom{-}0.05 & $211.195^{+0.610}_{-0.002}$ & 195.096 & 213.564 \\ 
\phantom{-}0.05 & -0.05 & $221.866^{+0.643}_{-0.002}$ & 202.986 & 224.175 \\ 
\phantom{-}0.10 & \phantom{-}0.10 & $202.608^{+0.388}_{-0.002}$ & 188.382 & 204.533 \\ 
\phantom{-}0.15 & \phantom{-}0.15 & $194.542^{+0.183}_{-0.001}$ & 182.347 & 196.652 \\ 
\phantom{-}0.15 & -0.15 & $221.887^{+0.637}_{-0.002}$ & 202.691 & 224.186 \\ 
\phantom{-}0.20 & -0.2 & $221.819^{+0.863}_{-0.003}$ & 202.698 & 224.201 \\ 
\phantom{-}0.20 & \phantom{-}0.20 & $187.838^{+0.020}_{-0.141}$ & 176.910 & 189.726 \\ 
\phantom{-}0.30 & \phantom{-}0.30 & $176.586^{+0.001}_{-0.653}$ & 167434 & 177.997 \\ 
\phantom{-}0.40 & -0.4 & $221.847^{+0.849}_{-0.003}$ & 202.771 & 224.357 \\ 
\phantom{-}0.40 & \phantom{-}0.00 & $188.138^{+0.008}_{-0.132}$ & 1776.927 & 189.759 \\ 
\phantom{-}0.40 & \phantom{-}0.40 & $167.545^{+0.002}_{-0.947}$ & 159.411 & 168.358 \\ 
\phantom{-}0.50 & -0.3 & $203.121^{+0.477}_{-0.002}$ & 188.440 & 204.657 \\ 
\phantom{-}0.60 & -0.6 & $222.080^{+0.808}_{-0.003}$ & 202.912 & 224.646 \\ 
\phantom{-}0.60 & \phantom{-}0.00 & $177.629^{+0.001}_{-0.645}$ & 167.462 & 178.053 \\ 
\phantom{-}0.60 & \phantom{-}0.60 & $154.139^{+0.005}_{-1.443}$ & 146.410 & 153.206 \\ 
\phantom{-}0.70 & -0.3 & $190.407^{+0.013}_{-0.164}$ & 177.005 & 189.915 \\ 
\phantom{-}0.70 & \phantom{-}0.30 & $160.935^{+0.004}_{-1.274}$ & 152.489 & 160.234 \\ 
\phantom{-}0.80 & -0.8 & $221.679^{+0.489}_{-0.002}$ & 203.145 & 225.114 \\ 
\phantom{-}0.80 & -0.5 & $198.993^{+0.237}_{-0.001}$ & 182.518 & 196.996 \\ 
\phantom{-}0.80 & \phantom{-}0.00 & $170.394^{+0.003}_{-1.026}$ & 159.445 & 168.433 \\ 
\phantom{-}0.80 & \phantom{-}0.20 & $162.069^{+0.005}_{-1.308}$ & 152.494 & 160.249 \\ 
\phantom{-}0.80 & \phantom{-}0.50 & $152.303^{+0.006}_{-1.640}$ & 143.642 & 150.049 \\ 
\phantom{-}0.80 & \phantom{-}0.80 & $145.357^{+0.006}_{-1.528}$ & 136.199 & 141.643 \\ 
    \hline
    \hline
    \end{tabular}
\end{table}

Finally, we compare the spin dependence of the NR scattering angles to PM predictions in  the small spin range. 
Using the simplifying fact that we are considering
equal-mass systems, and the relative scattering angle
$\theta_{\rm rel}=\frac12 (\theta_1+\theta_2)$, we can write, at quadratic order in spins, the simple general template
\begin{align}
\label{eq:spin_exp_eq}
\theta_{\rm rel}\left(\gamma,j;\chi_i\right) &\overset{(m_1=m_2)}{=} ~\theta_0\left(\gamma,j\right) + \, \theta_{\chi_+}\left(\gamma,j\right)  \chi_{+} \nonumber \\
&+ \theta_{\chi_{+}^2}\left(\gamma,j\right) \chi_{+}^2+ \theta_{\chi_{-}^2} \left(\gamma,j\right) \chi_{-}^2  + \mathcal{O}\left[\chi^3\right].
\end{align}

We can, in principle, determine the $\theta_i$ coefficients for the initial energy and angular momentum of our simulations, $(\hat{E}_{\rm in},\hat{L}_{\rm in}) \simeq (1.02264,1.1456)$ by fitting the scattering angles for configurations with small (absolute) values of the spins. 
We decided to focus on the 6 configurations with $|\chi_{1/2}| \leq 0.1$ and to fit them to the template Eq.~\eqref{eq:spin_exp_eq} (truncated to quadratic terms). 
We so obtained the following values:
\begin{align}
\label{eq:theta_NRcoefs}
\theta_0^{\rm fit} &= 3.872 \pm 0.048, \nonumber \\
\theta_{\chi_+}^{\rm fit} &= -4.211 \pm 0.031, \nonumber \\
\theta_{\chi_+^2}^{\rm fit} &= 8.75 \pm 0.60, \nonumber \\
\theta_{\chi_{-}^2}^{\rm fit} &= 0.6 \pm 3.1.
\end{align}
The reduced chi-squared of this fit is $\chi^2/(6-4) \simeq 1.17$ (corresponding to six data points and four degrees of freedom). This value is probably affected by the fact that we are assuming constant initial data, while the energy slightly changes between simulations. Let us emphasize that
we could not extract a meaningful value for the $\theta_{\chi_{-}^2}^{\rm fit}$ coefficient, consistently with our finding that these effects are subdominant (even for higher spin values).

In Table~\ref{tab:chiPM_spin} we compare these NR-fitting coefficients to their (PM-expanded and $w^{eob}$-resummed) analytical analogs.
In order to extract $w^{eob}_{\rm 3PM}$ and $w^{eob}_{\rm 4PM}$ (not including 4PM spin-orbit terms), we repeat the same fitting procedure employed for the NR angles (but setting the initial energy to the constant, average, value).
For the considered initial data, $(\hat{E}_{\rm in},\hat{L}_{\rm in}) \simeq (1.02264,1.1456)$, the PM-expanded predictions are not satisfactory, as could be deduced from Fig.~\ref{fig:chiPM_spin}.
The successive PM orders show a slow convergence towards the numerical results.
The EOB-resummed equivalents, instead, are in  good agreement with the numerical estimates, with the $w^{eob}_{\rm 4PM}$ values mostly agreeing within one standard deviation with the NR ones.

\begin{table}[t]
    \centering
        \caption{
        \label{tab:chiPM_spin}
        Small-spin dependence of the scattering angle for $\hat{E}_{\rm in} \simeq 1.02264$ and $\hat{L}_{\rm in}\simeq1.1456$. PM predictions for the different spin-orders (if available) together with values extracted by fitting NR data [Eq.~\eqref{eq:theta_NRcoefs}].
        }
    \setlength{\tabcolsep}{0.2cm}
    \begin{tabular}{c | c c | c c | c}
        \hline
        \hline
    & 3PM & 4PM & $w^{eob}_{\rm 3PM}$ & $w^{eob}_{\rm 4PM}$ & NR \\    \hline
    $\theta_0$            &  2.255     &  2.519  & 3.543 & 3.920 & $3.872 \pm 0.048$  \\
    $\theta_{\chi_{+}}$   & -0.433    & -0.621 & -2.895 & -4.221 & $-4.211 \pm 0.031$ \\
    $\theta_{\chi_{+}^2}$ &  0.085  & $\cdots$ & 3.898 & 7.904 & $8.75 \pm 0.60$ \\
    $\theta_{\chi_{-}^2}$ & -0.001 & $\cdots$ & 0.034 & 0.198 & $0.6 \pm 3.1$ \\
    \hline
    \hline
    \end{tabular}
\end{table}

\clearpage
\twocolumngrid
\bibliography{refs20231119.bib, local.bib}

\end{document}